\documentclass[useAMS,usenatbib]{mn2e}
\usepackage{graphicx}
\usepackage{amsmath}
\usepackage{amssymb}

\title[$N$-body Models of Rotating Globular Clusters]{$N$-body Models of Rotating Globular Clusters}
\author[A. Ernst, P. Glaschke, J. Fiestas, A. Just \& R. Spurzem]{A. Ernst$^{1,2}$\thanks{email:
aernst@ari.uni-heidelberg.de}, P. Glaschke$^{1,2}$, J. Fiestas$^{1}$, A. Just$^{1}$ and R. Spurzem$^{1}$\thanks{email: spurzem@ari.uni-heidelberg.de}\\
\\
$^{1}$Astronomisches Rechen-Institut / Zentr. Astron. Univ. Heidelberg, M\"onchhofstr. 12-14,
69120 Heidelberg, Germany\\
$^{2}$Max-Planck-Institut f\"ur Astronomie, K\"onigstuhl 17, 69117 Heidelberg, Germany}

\begin{document}

\date{Accepted ... Received ...}

\pagerange{\pageref{firstpage}--\pageref{lastpage}} \pubyear{2002}

\maketitle

\label{firstpage}

\begin{abstract}
We have studied the dynamical evolution of rotating globular clusters with direct $N$-body models. 
Our initial models are rotating King models; we obtained results for both equal-mass systems and systems composed out of two mass components. Previous investigations using a Fokker-Planck solver have revealed that rotation has a noticeable influence on stellar systems like globular clusters, 
which evolve by two-body relaxation. In particular, it accelerates their dynamical evolution through the gravogyro instability. 
We have validated the occurence of the gravogyro instability with direct $N$-body models. In the case of 
systems composed out of two mass components, mass segregation takes place, which competes with the rotation in the acceleration of the core collapse. The ``accelerating'' effect of rotation has not been detected in our isolated two-mass $N$-body models. Last, but not least, we have looked at
rotating $N$-body models in a tidal field within the tidal approximation. It turns out that rotation
increases the escape rate significantly. A difference between retrograde and
prograde rotating star clusters occurs with respect to the orbit of
the star cluster around the Galaxy, which is due to the presence of a ``third integral''
and chaotic scattering, respectively.
\end{abstract}

\begin{keywords}
$N$-body simulations -- rotating star clusters.
\end{keywords}

\raggedbottom

\section{Introduction}


This paper is devoted to an intriguing problem in the dynamics of globular clusters: 
How an overall rotation of a globular cluster
affects its dynamical evolution. 
There is a wealth of observational data which indicates that typical globular 
clusters are rotating. The first method for the
detection of rotation in globular clusters dates back in the first half of $20$th century. 
Observations showed that ``flattening'' (i.e. a deviation of spherical symmetry) is a common 
feature of globular clusters. 
The first ellipticity measurements have been done by H. Shapley in the 1920s, see Shapley (1930). 
Subsequent ellipticity measurments of Galactic and extragalactic globular clusters have 
been done, among others, by Geisler \& Hodge (1980), 
Frenk \& Fall (1982), Geyer et al. (1983), White \& Shawl (1987), Staneva, Spassova \& Golev (1996). 
The ellipticities of Galactic globular clusters are now given in the
Harris catalogue,\footnote{http://physwww.physics.mcmaster.ca/~harris/mwgc.dat}
cf. Harris (1996).
A second method for the detection of rotation of globular clusters relies on radial velocity
measurements of individual stars in globulars (Wilson \& Coffeen 1954,
Mayor et al. 1984, Meylan \& Mayor 1986, Gebhardt et al. 1994, 1995, 2000,
Reijns et al. 2006 and references therein).
The measured radial velocities can be plotted against the position angle. The amplitude of a 
sine curve which is fitted through the data points serves as a rough measure (i.e. a
lower limit) for the degree of rotation of the star cluster. 
Proper motions of individual stars have been measured as well and
used for the detection of rotation in globular clusters (e.g. Peterson \& Cudworth 1994,
van Leeuwen et al. 2000 or Anderson \& King 2003. The authors of the latter two references
conclude for $\omega$ Cen and 47 Tuc that their axes of rotation are considerably
inclined to the plane of the sky).
Table 7.2 in the review by Meylan \& Heggie (1997) gives already an overview of the 
degrees of rotation of Galactic globular clusters. For instance, Gebhardt et al. (1995) 
estimated $V_{\rm rot}/\sigma_{1D} \simeq 0.2$ for the evolved clusters M15 and 47 Tuc.
However, globular clusters in the Large and Small Magellanic Clouds are
younger and markedly more elliptical than typical Galactic globulars (see the above 
references), which suggests higher values of $V_{\rm rot}/\sigma_{1D}$.

From the theoretical side, the first published study of a rotating system of $N=100$ bodies 
is the work by Aarseth (1969) which aimed at the dynamical evolution of clusters of
galaxies. He already detected a flattening caused by rotation.
Subsequent attempts to study rotating globular clusters with direct $N$-body models 
have been made by Wielen (1974) for systems with particle numbers $N=50 - 250$
and Akiyama \& Sugimoto (1989) for $N = 1000$. Wielen concludes that 
``a slow overall rotation of a globular cluster does not significantly affect 
the dynamical evolution'', and Akiyama \& Sugimoto do not yet clearly demonstrate
numerically the existence of the gravogyro instability (see below) ``because of statistical 
noise due to a small number of particles''.
More recent numerical studies of rotating globular clusters have been done using the Fokker-Planck
[hereafter: FP] method, see the PhD thesis of Einsel (1996), Einsel \& Spurzem (1999) 
[hereafter ES99], the follow-up 
papers  by Kim et al. (2002, 2004) and the PhD thesis of Kim (2003) [hereafter K02, K03, K04].
Under certain approximations, the evolution of a phase space
distribution function which describes the macroscopic state of a globular cluster, 
is followed by the numerical solution of the FP equation. The advantage of such 
approximate models is their computational speed, i.e. one can do large parameter 
studies in a reasonable time. 
For the most recent FP models including rotation, which are 
compared directly to observations of M5, M15, 47 Tuc and $\omega$ Cen, see 
Fiestas et al. (2006).\footnote{A database of rotating star cluster models exists at the ARI:
http://www.ari.uni-heidelberg.de/clusterdata/}

The present study is again concerned with $N$-body models of rotating globular clusters.
Based on Newton's law of gravitation, $N$-body modelling is probably the most direct way to 
investigate the dynamics of stars in globular clusters. However, it is computationally expensive
and time-consuming.\footnote{Therefore $N$-body codes for massively parallel compšuters have 
been written. Moreover, special-purpose hardware systems like GRAPE boards have been in 
use for $N$-body simulations since the early 1990s (see http://grape.astron.s.u-tokyo.ac.jp/grape/). 
A new development is a technology based on FPGAs (Field Programmable Gate Arrays).
However, such hardware is not used in the present study.}
On the other hand, direct $N$-body modelling particularly allows
to test the validity of assumptions used in statistical modelling of globular clusters, like 
FP models. 

This paper is organized as follows:
Section 2 describes our initial models and the numerical method. In 
Section 3 we will have a closer look at the effect of rotation on the core collapse in isolated 
equal-mass models. It is well-known that the core collapse is caused by the gravothermal instability:
The negative heat capacity of the self-gravitating core of the star cluster leads to an amplification
of temperature gradients if the halo of the star cluster is extensive enough.
The core collapse is stopped by the formation of a few binaries in the core of the star cluster
by three-body processes (Giersz \& Heggie 1994b, 1996). These produce heat at a 
rate slightly higher than the rate of energy loss due to two-body relaxation, 
thereby cooling the cluster core which has negative heat capacity.
If the star cluster rotates, we numerically demonstrate in Section 3 that an additional
instability occurs: The gravogyro instability, which accelerates the core collapse
in our equal-mass models.
In Section 4, which is dedicated to tidally limited models, we additionally study the escape of 
stars from rotating star clusters, i.e. mass loss\footnote{The term ``mass loss'' is not used in the 
usual sense (due to stellar evolution), but in the sense of a mass loss of the whole star cluster 
through escaping stars (i.e. due to dynamical interactions).} across the tidal 
boundary. In Section 5, which treats rotating systems composed out of two
mass components, we study the interplay between rotation and mass segregation.

\section{Initial models / Numerical method}

We employ generalized King models with rotation. The distribution function
is given by

\begin{eqnarray}
f(E,J_z)=C\cdot
\left[\exp\left(-\frac{E-\Phi_t}{\sigma_K^2}\right)-1\right]\cdot
\exp\left(-\frac{\Omega_0 J_z}{\sigma_K^2}\right), \label{eq:rotking}
\end{eqnarray}

\noindent
which depends on the integrals of motion $E$ (energy) and $J_z$
($z$-component of angular momentum),
where $\Phi_t$ is the potential at the outer boundary of the model
(i.e. at the radius where the density approaches zero),  $\sigma_K$ is the King velocity dispersion
(which is not the central velocity dispersion)
and $\Omega_0$ is close to the angular speed in the cluster center (which has
nearly solid body rotation). 
In the distribution function (\ref{eq:rotking}),
the first two factors on the right-hand side represent just a usual King model.
The last factor, which depends on $J_z$, introduces rotation into the model, i.e.
a difference between positive and negative $J_z$. It also allows for
anisotropy, i.e. we have $\sigma_r = \sigma_z \not= \sigma_\phi$ for the
velocity dispersions in radial, vertical and tangential directions, respectively.
Note that models where $f \propto \exp(-\gamma J_z)$ have been first
used by Prendergast \& Tomer (1970)  and Wilson (1975) to model 
elliptical galaxies. However, this term in combination with a King model
has been first used in Goodman (1983, unpublished), then by Lagoute \&
Longaretti (1996), Longaretti \& Lagoute (1996), Einsel (1996), ES99, K02, K03 and K04,
Fiestas (2006) and Fiestas et al. (2006).

The rotating King model based on the distribution function (\ref{eq:rotking}) 
possesses rotational symmetry and differential rotation.
It has two dimensionless free parameters: The
King parameter $W_0 = - (\Phi_0 - \Phi_{\rm t})/\sigma^2_{\mathrm{K}}$,
where $\Phi_0$ is the central potential, and the rotation parameter 
$\omega_0 = \sqrt{9/(4 \pi G \rho_c)} \cdot \Omega_0$,
where $G$ is the gravitational constant and $\rho_c$ is the central
density. It reduces to a King model in the limit $\omega_0=0$. 
ES99 give the ratio $\lambda = T_{\rm rot}/T_{\rm kin}$ of rotational
to kinetic energy in percent in their Table 1 for rotating King models with $W_0=6$. 
The ratio of the root mean squared 
rotational velocity to the velocity dispersion is then given by
$V_{\rm rot}/\sigma_{1D} = \sqrt{3\lambda/(1-\lambda)}$, which may be compared
with observations. Furthermore, it may be of interest to note that between 
$0.7 < \omega_0 < 0.8$ there is a transition where the initial axisymmetric 
isolated rotating King models become unstable to the formation of a bar 
(P. Berczik, priv. comm.).

Our models composed out of two mass components are initially uniformly mixed
and characterized by two additional dimensionless parameters, the stellar mass 
ratio $\mu=m_2/m_1$, where $m_1$ and $m_2$ are the individual masses of the 
light and heavy stars, respectively, and the mass fraction of heavy stars $q=M_2/M$, 
where $M_2$ is the total mass of the heavy component and $M$ is the total mass
of the star cluster.

In direct $N$-body models, the main constraint for a reasonable computing time 
is the number of particles $N$. In our simulations, we use $N=5000$ for all
models with $\mu \leq 10$ (including all equal-mass models)
and $N=32000$ for models with $\mu=25$ and $\mu=50$.
A typical simulation for one of our equal-mass models took about three
days on a 3 GHz Pentium PC, until the core collapse time was reached.

The codes used are {\sc nbody6++} (e.g. Spurzem 1999) and {\sc nbody6}
(see Aarseth 1999, 2003 for an overview). 
The former code is a variant of the latter modified in order to run on 
massively parallel computers. Both codes share, aside from parallelization,
the same fundamental features and yield comparable results. A fourth-order Hermite
scheme, applied first by Makino \& Aarseth (1992), is used for the 
direct integration of the Newtonian equations of motion of the $N$-body system. 
The codes use
adaptive and individual time steps, which are organized in hierarchical 
block time steps, the Ahmad-Cohen neighbor scheme (Ahmad \& Cohen 1973),
Kustaanheimo-Stiefel regularization of close encounters
(Kustaanheimo \& Stiefel 1965) and 
Chain regularization (Mikkola \& Aarseth 1990, 1993, 1996,1998). 

All models done with {\sc nbody6++} have been calculated on single-processor 
machines at the ARI, except the two-mass models TM9-TM12 (see section 5).
Some models (from series EM2, see Section 3) have been calculated using Aarseth's
original code {\sc nbody6}.
For some models, we used averaging over several runs with the same particle
number $N$ but different initial values for positions and velocities. 
This was done in order to obtain a better statistical quality, cf. Section 2.2 of Giersz 
\& Heggie (1994a) for a discussion of this approach. Averaging increases the 
computing time for gravitational forces of a system of $N$ bodies on a linear scale, 
whereas increasing the particle number $N$ increases the computing time for 
gravitational forces $\propto N^2$ in the limit of high $N$. 
Since we consider in the present work processes acting on the relaxation time scale,
an extra power of $N$ contributes to the scaling of the total computing time of a model.
In our averaged models, the physical quantities shown in the plots 
are determined as the arithmetic mean of those values of the considered quantity, 
which resulted from the individual runs.

The quantities on the vertical axis of our figures are shown in dimensionless $N$-body units, 
i.e. $G=M=-4E=1$, where $G$ is the gravitational constant, $M$ is the total
mass of the system and $E$ the total energy, see Heggie \& Mathieu (1986).
The resulting length unit is the virial radius $r_V=GM^2/(-4E)$, which is of the
order of the half-mass radius of a globular cluster, see, for instance, Table 1 in G\"urkan et al. (2004). 
The resulting time unit is then given by $t_V = (GM/r_V^3)^{-1/2}$.
The time on the horizontal axis of most figures is shown in units of the 
initial half-mass relaxation time, i.e. the $t_{\rm rh}$ means $t_{\rm rh}(0)$.
For the half-mass relaxation time, we adopted 
the expression

\begin{equation}
t_{\rm rh} = \frac{8\pi}{3} \frac{\left[\sqrt{2\vert E\vert/M} \, r_V\right]^3 N}{15.4 \, G^2 \, M^2 \, \ln(\gamma N)}  \label{eq:trh2}.
\end{equation}

\noindent
with $\gamma=0.11$. The above definition is 
based on Equation (5) of the paper by Spitzer \& Hart (1971) with

\begin{equation}
n = \frac{3 N}{8\pi r_V^3}, \ \ \ \ \ v_m = \sqrt{\frac{2\vert E\vert}{M}}, \ \ \ \ \ m = \frac{M}{N}
\end{equation}

\noindent
for the mean value of the particle density inside $r_V$, the root mean squared stellar velocity
(i.e. the 3D velocity dispersion) and the mean stellar mass, respectively. 
It may be of interest to note that (\ref{eq:trh2}) can be written in terms of $t_V$ as

\begin{equation}
t_{\rm rh} = \frac{2\sqrt{2} \, \pi N}{3\cdot 15.4 \ln(\gamma N)} \, t_V \label{eq:trh}.
\end{equation}

\noindent
For $(N=5000, \gamma = 0.11)$ we have $t_{\rm rh}\simeq 152 \, t_V$.The crossing time at the virial radius is given by $t_{cr} =2r_V/v_m = 2\sqrt{2} \, t_V$.





For the data evaluation, we employed boxcar smoothing to reduce 
noise from our plots. We usually employed smoothing widths of $5 - 20 \, t_V$.

\section{Isolated equal-mass models}

\medskip
\begin{table}
\begin{minipage}{\textwidth}
\begin{tabular}{l|l|llc|l|l}
\hline
\hline
Model & $W_0$ & $\omega_0$ & N & Averaging & $t_\mathrm{cc}/t_\mathrm{rh}$ \\
\hline
EM1a & 3 & 0.0 & 5K & no & 10.68 \\
EM1b & 3 & 0.3 & 5K & no & 10.10 \\
EM1c & 3 & 0.6 & 5K & no & 10.35 \\
\hline
EM2a & 6 & 0.0 & 5K & 4 & 7.51 $\pm$ 0.52 \\
EM2b & 6 & 0.3 & 5K & 4 & 7.45 $\pm$ 0.25 \\
EM2c & 6 & 0.6 & 5K & 4 & 7.03 $\pm$ 0.19\\
EM2d & 6 & 0.9 & 5K & 3 & 6.19 $\pm$ 0.35\\
\hline
\end{tabular}
\end{minipage}
\caption{The initial isolated equal-mass models. For the models of series EM2 we also
give the run-to-run variation $\sigma_{n-1}$ in $t_{cc}/t_{rh}$ in the last column.}
\label{EM12}
\end{table}

\medskip 

In this section, we study the effect of rotation on the core collapse in isolated $N$-body
models (i.e. without tidal field) and find, that these models suggest the occurence of the 
gravogyro instability, which has been found earlier in FP models. 
Table \ref{EM12} shows our initial isolated equal-mass models. 


\subsection{Core collapse}

For the theory of the gravothermal instability, which leads to the
core collapse of star clusters we refer to the original works of
Antonov (1962), Lynden-Bell \& Wood (1968) and Hachisu et al. (1978a/b). 
It is nowadays also described in many textbooks such as 
Binney \& Tremaine (1987) or Heggie \& Hut (2003).

\begin{figure}
\includegraphics[width=0.45\textwidth]{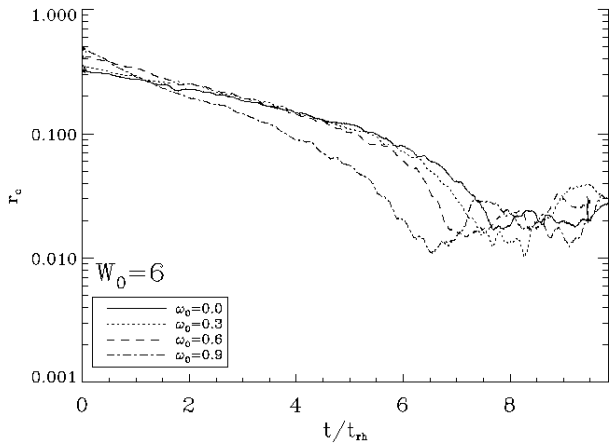} 
\caption{Time evolution of the core radius for the isolated
models EM2a-d.} 
\label{fig:rcore2}
\end{figure}

Figure \ref{fig:rcore2} shows the time evolution of the core radius for the 
models of series EM2. In this series, one very fast rotating model 
with $\omega_0=0.9$ is included. The core radius is defined as

\begin{equation}
r_c=\sqrt{ \frac{\sum_{i=1}^N \vert \vec r_i- \vec r_{d} \vert^2 \rho_{i}^2}{\sum_{i=1}^N \rho_{i}^2} },
\end{equation}

\noindent
where $\vec{r}_i$ is the position of the $i$th star, $\vec{r}_d$ are
the coordinates of the density center of the star cluster 
and $\rho_i$ is the local density at the position of the $i$th star.
The latter two quantities are calculated according to the
description in Casertano \& Hut (1985) using the distance to the
fifth nearest neighbor.  It can be seen in Figure \ref{fig:rcore2} that the core radius 
decreases by two orders of magnitude and that the core collapse occurs earlier 
for the faster rotating models. Note that the four models of series EM2 do not 
exactly have the same initial core radius in the beginning, i.e. the core radii differ by factors 
up to approximately two. This indicates that the models are indeed 
not in all respects exactly comparable, cf. Figures 14/15 of K02. 
They have slightly different density distributions corresponding to
their degree of flattening.
However, since the core shrinks by two orders of magnitude during 
the core collapse, a factor of two in the difference of the initial core 
radii seems to be negligible.

\begin{figure*}
\includegraphics[width=0.85\textwidth]{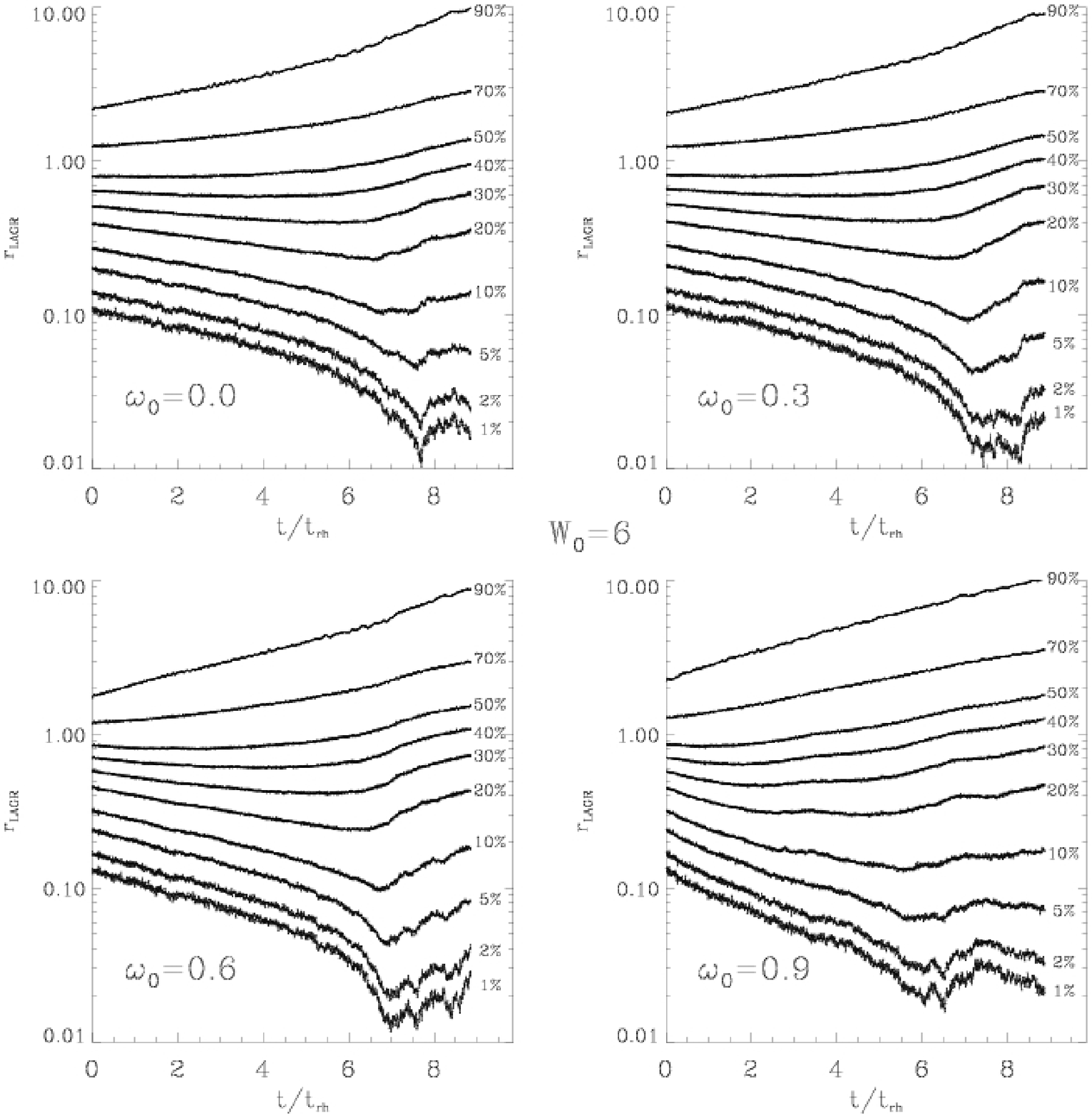} 
\caption{Time evolution of the Lagrangian radii, which contain 
1\%, 2\%, 5\%, 10\%, 20\%, 30\%, 40\%, 50\%, 70\% and 90\% of the initial total mass 
of the star cluster for the isolated equal-mass models EM2a-d. The inner
Lagrangian radii contract, while the outer ones expand in the core
collapse phase.} 
\label{fig:rlagr}
\end{figure*}

Figure \ref{fig:rlagr} shows the time evolution of the Lagrangian radii
for the models of series EM2. 
The inner Lagrangian radii shrink and the outer ones expand.

The last column of Table \ref{EM12} contains the core collapse times for the 
isolated equal-mass models. The core collapse time is defined as the time when 
the core radius $r_c$ reached its first sharp minimum. For the averaged runs
the given time is the arithmetic mean of the core collapse times which we 
determined for each single run. We also give the standard deviations $\sigma_{n-1}$
as a measure of the run-to-run variation of the core collapse times.

\subsection{Gravogyro instability revisited - I}

\begin{table}
\begin{minipage}{\textwidth}
\begin{tabular}{cc}
\hline
\hline
Gravothermal instability & Gravogyro instability \\
\hline
$\partial_r \sigma^2 \not= 0$ & $\partial_r \omega \not= 0$ \\
Negative specific heat & Negative specific moment of \\
 capacity & inertia \\
Energy transport & Angular momentum transport \\
Heat conduction & Viscosity \\
\hline
\end{tabular}
\end{minipage}
\caption{A comparison between gravothermal and gravogyro instability.}
\label{comparison}
\end{table}

There is obviously a need to explain the accelerated core collapse of the
rotating models in Section 3.1. A theoretical model, which
explains such an effect, has been proposed as early as in the 1970s 
by Inagaki \& Hachisu (1978) and Hachisu (1979, 1982). They called
it the gravogyro instability. This instability is expected in self-gravitating and
rotating systems and has been derived for self-gravitating cylinders of infinite
length in $z$-direction: If we equate the centrifugal force to the gravity in the
equatorial plane ($z=0, R^2=x^2+y^2$) and neglect the pressure gradient,
we have

\begin{equation}
\frac{2GM_R}{R} \simeq \frac{j_z^2}{R^3},
\end{equation}

\noindent
where $M_R$ is the mass per unit length in
$z$-direction contained within the radius $R$ in cylindrical coordinates and $j_z$ 
is the $z$-component of specific angular momentum. We readily obtain
$\delta j_z/j_z \simeq \delta R/R$. 
Using the equation $j_z = R^2\omega$ in addition, we obtain
$\delta j_z \simeq \omega R \delta R$ and
$\delta j_z = 2\omega R \delta R + R^2\delta\omega$. The combination of
these relations yields the following linear differential relation between specific 
angular momentum $j_z$ and angular speed $\omega$: 

\begin{equation}
\delta j_z \simeq -R^2 \delta \omega. \label{eq:gg2}
\end{equation}

\noindent
A negative specific moment of inertia occurs in this relation, cf. Inagaki \& 
Hachisu (1978).


If we let an initial model evolve, which exhibits a radial gradient of the angular speed 
and has reasonable rotation curve, 
angular momentum is transported outwards on the relaxation time scale by the 
stellar dynamical analog of viscosity. The core of the star cluster  contracts because of 
a deficit in the centrifugal force. The angular speed in the core increases according 
to relation (\ref{eq:gg2}). A runaway departure from the initial state takes
place. This cycle is what they denoted as the gravogyro instability. For a comparison
between gravothermal and gravogyro instabilities see Table \ref{comparison}.

\begin{figure}
\includegraphics[angle=90,width=0.45\textwidth]{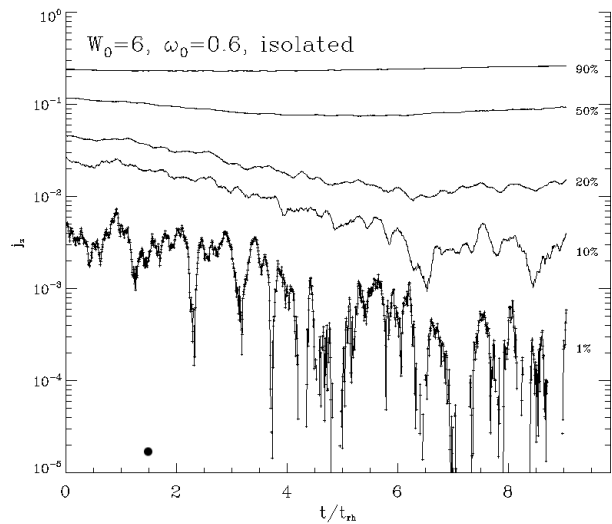}
\caption{Time evolution of the $z$-component of the specific angular 
momentum of the particles within the Lagrangian radii for the fast
rotating isolated model EM2c. The isolated points are artefacts and stem (like
the interruptions in the 1\% curve) from the sign change of $j_z$.}
\label{fig:jspec1}
\end{figure}

\begin{figure*}
\includegraphics[angle=90,width=0.85\textwidth]{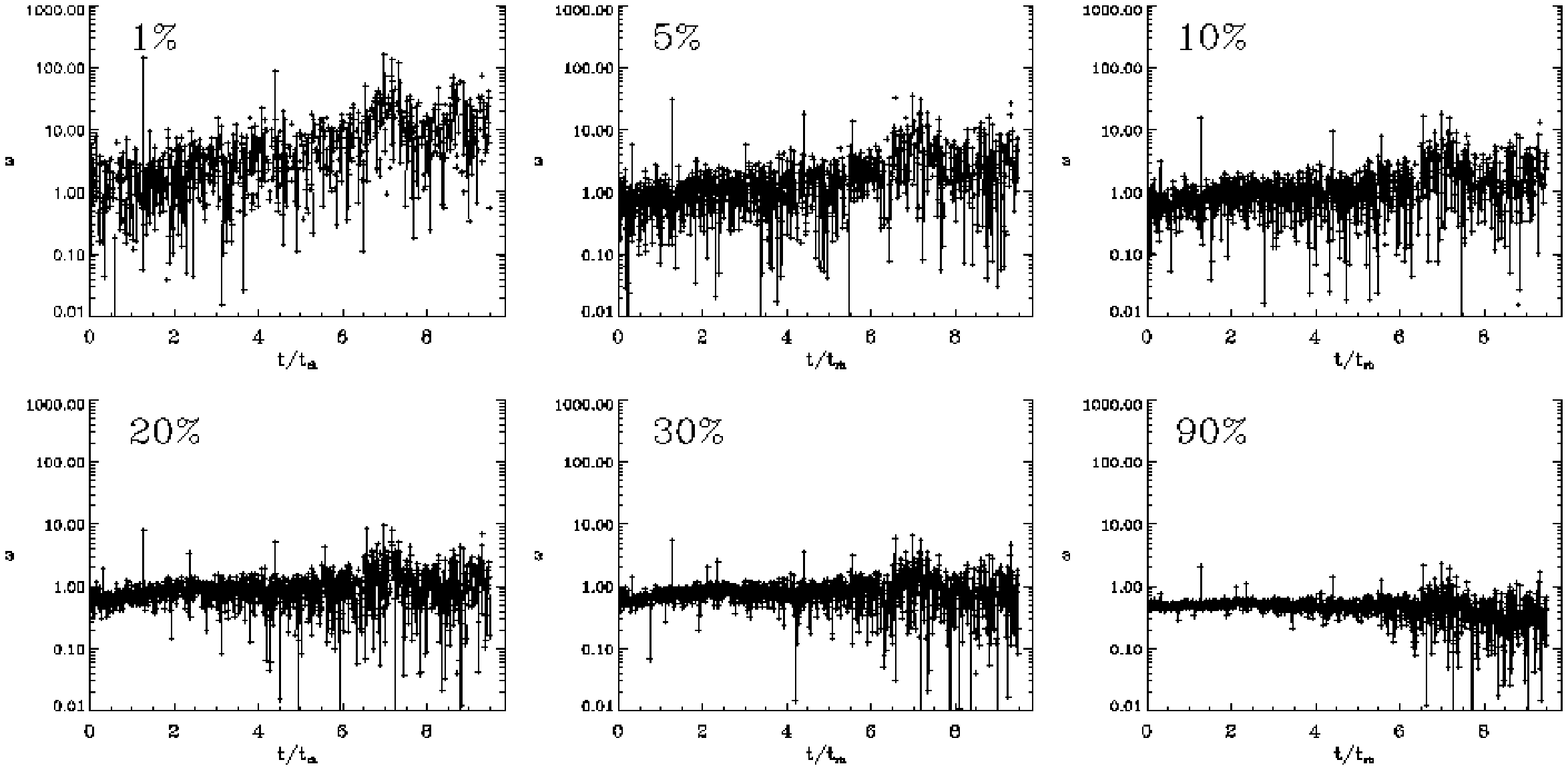}
\caption{Time evolution of the average angular speed of the particles within 
the Lagrangian radii for the fast rotating isolated model EM2c with $W_0=6, \omega_0=0.6$.
The interruptions and isolated points are artefacts stemming from the sign change of $\omega$.}
\label{fig:omega1}
\end{figure*}

One may wonder: Does the gravogyro catastrophe really occur in isolated 
$N$-body models? Figure \ref{fig:jspec1} shows the time evolution of the $z$-component of 
specific angular momentum within the Lagrangian radii for the fast rotating isolated model EM2c. 
The $z$-component of specific angular momentum of each star is calculated according to

\begin{equation}
j_\mathrm{z}= [(\vec{r}_\star-\vec{r}_d) \times \vec{v}_\star \, ]_z, \label{eq:jz1}
\end{equation}

\noindent
where $\vec{r}_\star=(x_\star,y_\star,z_\star)$ is the position of a particle, 
$\vec{v}_\star=(v_{\star x},v_{\star y},v_{\star z})$ is its 
velocity and $\vec{r}_d=(x_d,y_d,z_d)$ is the 
density center of the star cluster.
It is then summed over all particles within a given Lagrangian radius and divided by
the number of particles inside that radius.
One notes that angular momentum does indeed diffuse from the inner parts of 
the cluster to the outer parts, as time proceeds.

\noindent
Now we study the evolution of the angular speed $\omega$.
The gravogyro instability implies that 
$j_\mathrm{z}$ goes down while $\omega$ increases.
Figure \ref{fig:omega1} shows the time evolution of the average angular speed of
the particles within the Lagrangian radii. 
The angular speed of one particle is calculated according to

\begin{equation}
\omega=\frac{j_\mathrm{z}}{R_\star^2}
= \frac{[(\vec{r}_\star-\vec{r}_d)\times\vec{v}_\star \, ]_z}{R_\star^2},
\end{equation}

\noindent
with the same notations as in equation (\ref{eq:jz1}), but 
$R_\star=\sqrt{(x_\star-x_d)^2+(y_\star-y_d)^2}$ is the radius with respect to
the density center in cylindrical coordinates. 
It is then summed over all particles inside a given Lagrangian radius and divided 
by the number of particles within that radius.
One can observe an increasing average angular speed inside those Lagrangian 
radii, which show a decrease in the $z$-component of angular momentum. Thus, it is 
shown that gravogyro effects appear in our isolated models. 

\section{Tidally limited equal-mass models}

\begin{figure*}
\includegraphics[angle=90,width=1.0\textwidth]{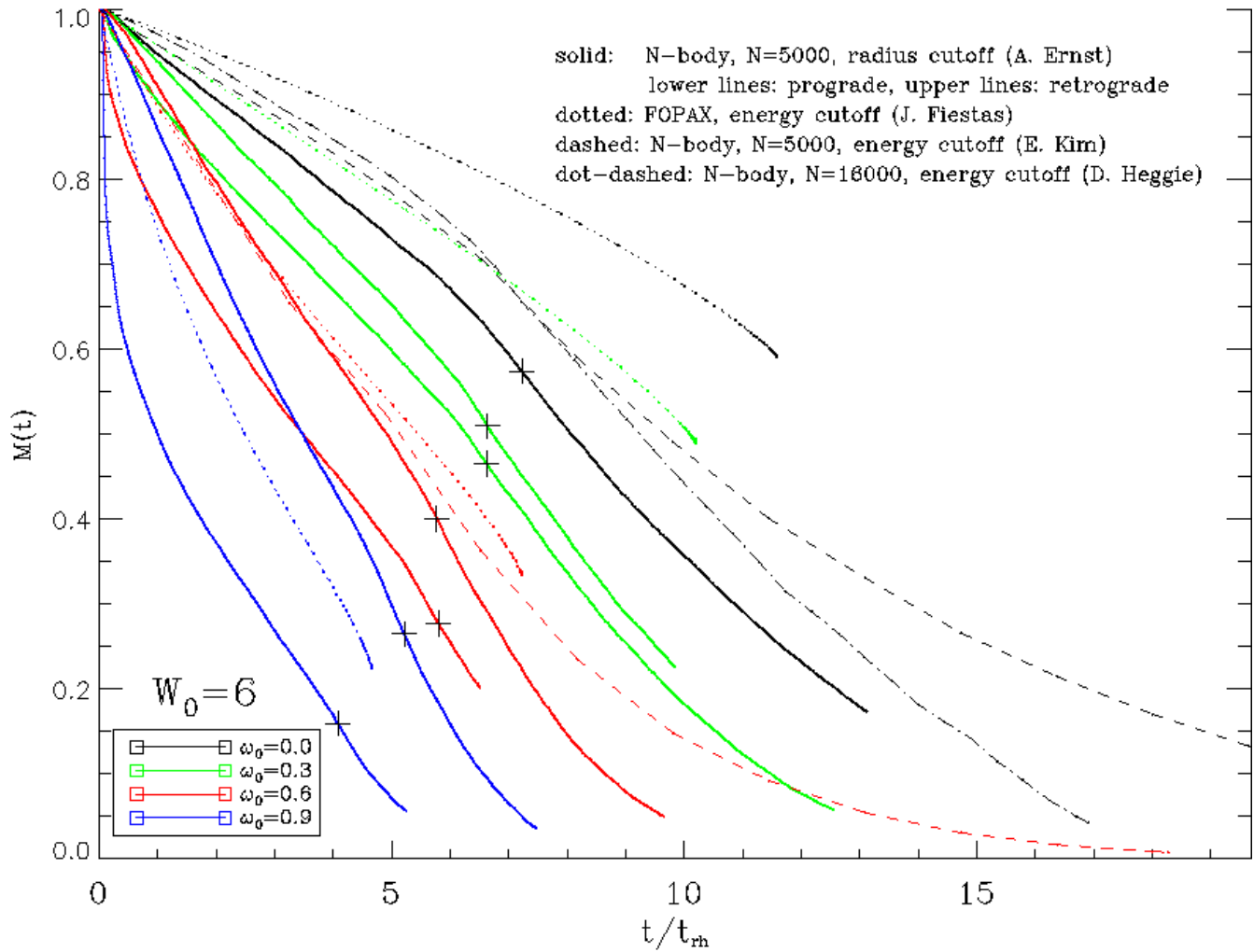}
\caption{Time evolution of the total mass of the star clusters for the prograde
rotating tidally limited models EM3a-d (lower solid lines), the retrograde
rotating tidally limited models EM4b-c (upper solid lines), FP models with energy 
cutoff (dotted), two $N$-body models with $N=5000$ and 
energy cutoff (dashed) and one non-rotating $N$-body model 
with $N=16000$ and energy cutoff (dot-dashed). 
The core collapse times of our $N$-body models are marked with a
cross.
}
\label{fig:massloss}
\end{figure*}

As a step towards more realistic models of globular clusters, we 
investigate in this section the effects of the tidal field of the Galaxy
on the dynamical evolution of rotating globular clusters.
The implementation of the tidal field of the Milky Way Galaxy within
the tidal approximation used in {\sc nbody6} and {\sc nbody6++} is in 
detail described in Appendix A. Different escape criteria are
discussed in Appendix B. In our models, the star cluster initially
completely fills its Roche lobe in the tidal field of the Galaxy.
In other words, the density of the initial rotating King model 
approaches zero at the physical tidal cutoff radius.
The tidal radius is defined as the distance of the star cluster 
center to the Lagrangian points $L_1 / L_2$.

Table \ref{EM34} shows the initial tidally limited equal-mass models. The model of series EM3
are prograde rotating, i.e. the stars move around the $z$-axis of the cluster-centered coordinate 
system in the same sense as the star cluster moves around the Galaxy. The models of series 
EM5 are retrograde rotating, which we denoted by a negative sign of the rotation
parameter $\omega_0$ for clarity. The only difference between the rotating initial models 
of series EM3 and EM4 is that we reversed the sign of all initial velocity vectors.

\begin{table}
\begin{minipage}{\textwidth}
\begin{tabular}{l|lrlcl|l|l|l}
\hline
\hline
Model & $W_0$ & $\omega_0$ & N & Averaging & $t_\mathrm{cc}/t_\mathrm{rh}$\\
\hline
EM3a & 6 & 0.0 & 5K & 4 & 7.11 $\pm$ 0.37\\
EM3b & 6 & 0.3 & 5K & 4 & 6.37 $\pm$ 0.36 \\
EM3c & 6 & 0.6 & 5K & 4 & 5.42 $\pm$ 0.33 \\
EM3d & 6 & 0.9 & 5K & 4 & 3.97 $\pm$ 0.40 \\
\hline
EM4b & 6 & -0.3 & 5K & 4 & 6.50 $\pm$ 0.33 \\
EM4c & 6 & -0.6 & 5K & 4 & 5.62 $\pm$ 0.07 \\
EM4d & 6 & -0.9 & 5K & 4 & 4.88 $\pm$ 0.28 \\
\hline
\end{tabular}
\end{minipage}
\caption{The initial tidally limited equal-mass models. The
rotating models of series EM3 are prograde rotating while the models of series
EM4 are retrograde rotating with respect to the orbit of the star cluster around
the Galaxy. We also give the run-to-run variation $\sigma_{n-1}$ in $t_{cc}/t_{rh}$ 
in the last column.}
\label{EM34}
\end{table}

\subsection{Core collapse}

\begin{figure}
\includegraphics[width=0.45\textwidth]{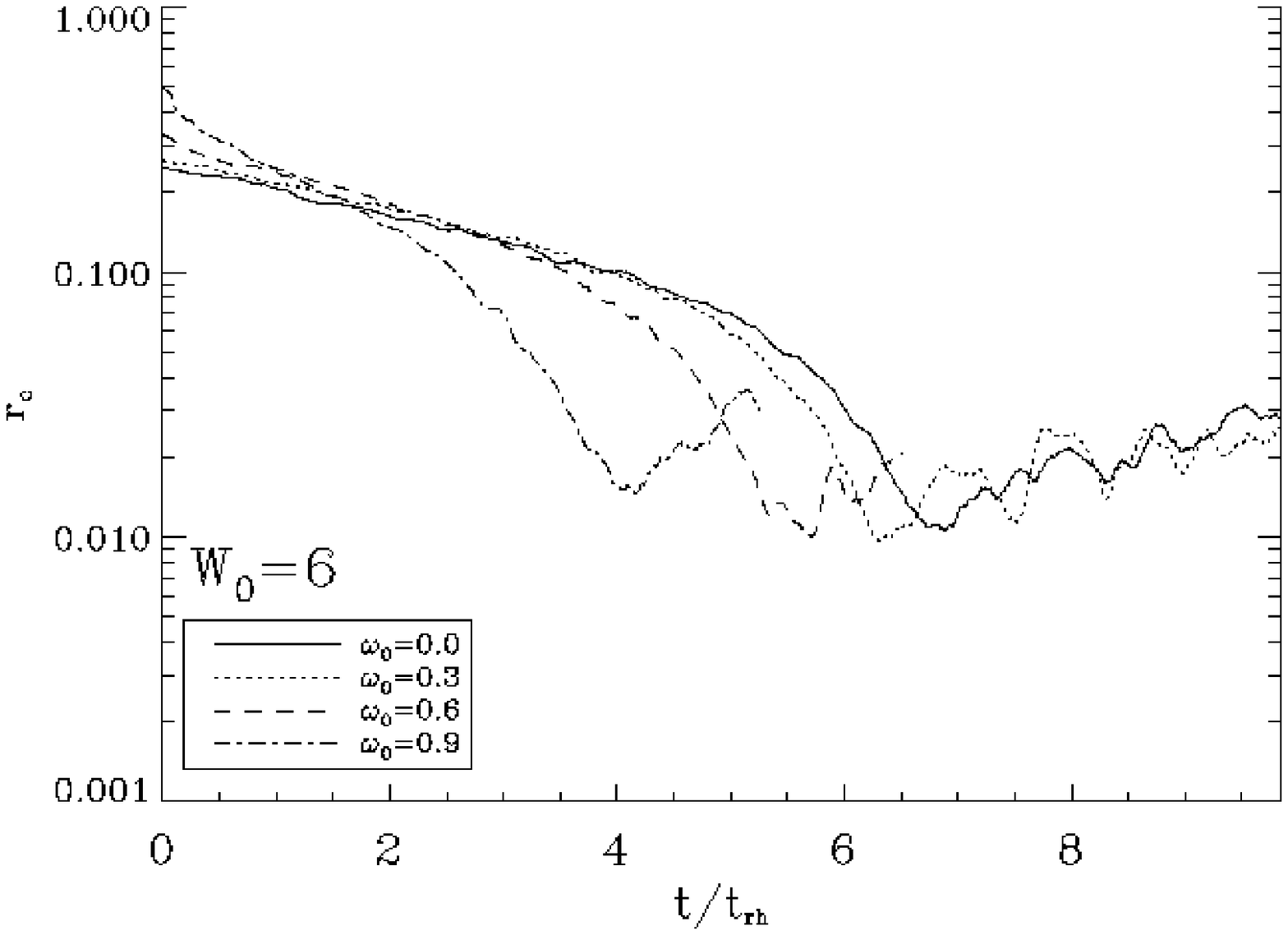} 
\caption{Time evolution of the core radius for the prograde rotating tidally limited
models EM3a-d} 
\label{fig:rcore3}
\end{figure}

Figure \ref{fig:rcore3} shows the time evolution  of the core radius for
the prograde rotating tidally limited models EM3a-d.
A comparison of Figures \ref{fig:rcore2} and \ref{fig:rcore3} and the core collapse times in 
Tables \ref{EM12} and \ref{EM34} shows, that the tidal field
further accelerates the core collapse by a significant amount. 
The reason for this behaviour is that rapid mass loss across the tidal boundary 
provides an additional mechanism for the 
transport and removal of energy and angular momentum, thus enhancing the effects of
the gravothermal and gravogyro instabilities.  
An inspection of Table \ref{EM34} reveals that there is a difference between the core collapse 
times between the models of series EM3 and EM4, i.e. the prograde rotating models of
series EM3 collapse faster than the retrograde rotating model of series EM4. By considering 
the run-to-run variation $\sigma_{n-1}$ in the last column of table \ref{EM34}, we see that the 
statistical significance of this effect is strongest for the two models with 
$\omega_0=0.9$
as compared with the case of the slower rotating models..
Therefore this effect might be related to the formation
of a bar in the fastest rotating models. On the other hand, the difference of the core collapse
times between the prograde and retrograde rotating models may also be related 
to the difference in the rates of mass  loss (see next Section): The loss of energy and
angular momentum through escaping stars is more rapid for the prograde rotating models and
 thus the effects of the gravothermal and gravogyro instabilities, which 
accelerate the core collapse, are stronger.

\subsection{Mass loss}

Figure \ref{fig:massloss} shows the time evolution of the total mass of the globular clusters, 
where the initial mass is normalized to one, for several models,
including simulations with the 2D FP code {\sc fopax} (with energy cutoff, cf. Fiestas 2006), 
one rotating and one non-rotating 
$N$-body model by E. Kim (with energy cutoff, without the tidal approximation, 
see K03) and one non-rotating $N$-body model by D. C. Heggie (with energy cutoff,
without the tidal approximation). 
The solid lines are our {\sc nbody6++} results (with radius cutoff and the tidal
approximation).
The different ``cutoffs'' are explained in Appendix B. 
The initial models are the same rotating (or non-rotating) 
King models. Note that for the rotating models, there are
two solid lines for each color. The lower of these lines corresponds
to a prograde rotating model (i.e., from series EM3). The upper of these lines 
corresponds to a retrograde rotating model (i.e., from series EM4). The escape rate for 
retrograde rotating models is considerably lower, since many stars in retrograde 
orbits are subject to a ``third integral'' of motion, which restricts their 
accessible phase space and hinders them from escaping, see 
Fukushige \& Heggie (2000) and Appendix B. 

Figure \ref{fig:massloss} also shows, that the faster rotating models suffer 
stronger mass loss than the slowly or non-rotating models, a
fact which has already been shown in the FP models of ES99.

The differences between the models of the same rotation parameter
are due to different implementations of the tidal field in
the various codes (cf. Appendix B) or due to the different 
modelling techniques in general. 
For instance, the difference between the $N$-body and FP models is 
the tidal approximation used in {\sc nbody6} and {\sc nbody6++}, 
i.e. the modified equations of motion with a linear approximation of the 
tidal forces (see Appendix A), a feature, which cannot be implemented 
in an FP code, which relies only only on a tidal cutoff. 
The geometry of the almond-shaped tidal boundary within the tidal 
approximation (see Figure \ref{fig:hillsurf}) differs from the geometry 
of our star cluster models, which are axisymmetric. This may cause
another difference between $N$-body and FP models.
In addition, the difference between prograde and retrograde
rotating models cannot be seen in FP models.
Note that Kim et al. (2006) use for their comparison $N$-body
models with an artificially implemented tidal cutoff and no
modification of the equations of motion.

Last but not least, a further inspection of Figure \ref{fig:massloss} reveals,
that core collapse causes an increase in the escape rate, i.e.
short before the core collapse time the curves steepen slightly.
The moment of core collapse is marked with a cross on each curve corresponding
to one of our $N$-body models. However, our main aim was to show the difference of 
prograde and retrograde rotating models and the fact, that rotation significantly
increases the rate of mass loss.

\subsection{Gravogyro instability revisited - II}

\begin{figure}
\centering
\includegraphics[angle=90,width=0.45\textwidth]{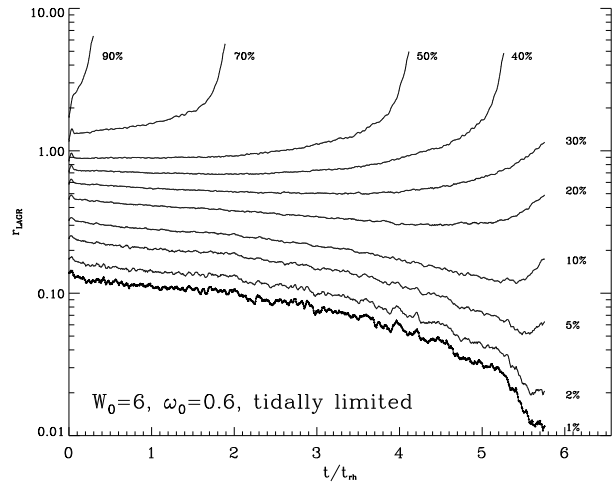}
\caption{Time evolution of the Lagrangian radii for a single run of 
the fast rotating tidally limited model EM3c. For explanations see the text.}
\label{fig:figrlagr}
\end{figure}

\begin{figure}
\centering
\includegraphics[angle=90,width=0.45\textwidth]{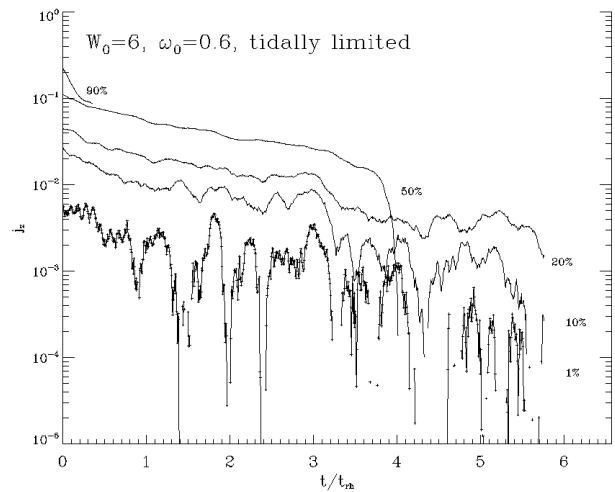}
\caption{Time evolution of the $z$-component of the specific angular 
momentum of the particles within the Lagrangian radii for a single run of 
the fast rotating tidally limited model EM3c. For explanations see the text.}
\label{fig:jspec2}
\end{figure}

\begin{figure*}
\centering
\includegraphics[angle=90,width=0.85\textwidth]{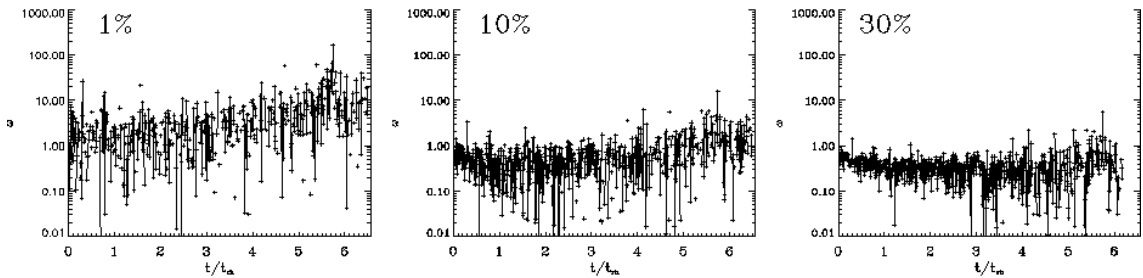}
\caption{Time evolution of the average angular speed of the particles within the Lagrangian 
radii for a single run of the fast rotating tidally limited model EM3c with $W_0=6, \, \omega_0=0.6$. 
For explanations see the text.}
\label{fig:omega2}
\end{figure*}

In this subsection, we briefly look at the gravogyro instability in the tidally 
limited $N$-body models. The mass loss through the tidal boundary provides, as
noted before, an additional mechanism for the transport and removal of
angular momentum (besides viscosity effects).
Figure \ref{fig:figrlagr} shows the time evolution of the Lagrangian radii for one 
run of the tidally limited model EM3c. The Lagrangian radii are defined with
respect to the initial particle number. Therefore the particle number within
the Lagrangian radii remains constant while the system evolves, whereas after 
some time, the outer Lagrangian shells are completely lost due to escaping stars, i.e. 
the curves for the outer Lagrangian radii bend upwards until they reach twice the tidal radius, 
which is the cutoff radius in the adopted escape criterion.
Figure \ref{fig:jspec2} shows the time evolution of the $z$-component of
specific angular momentum within the Lagrangian radii for the same single run of the tidally 
limited model EM3c. The curves for $j_z$ decrease due to escaping stars
carrying away angular momentum. The curves stop at the point at which
the adjacent outer Lagrangian shell is lost, such that we always have a constant
particle number within the Lagrangian radii.
Figure \ref{fig:omega2} shows the time evolution of the average angular speed of 
the stars within the Lagrangian radii for the same single run 
of the tidally limited model EM3c. 
After an initial decrease of $\omega$ due to mass loss, which is most prominent in the
outer Lagrangian spheres but reaches deep into the core, $\omega$ increases
more than in the isolated model and up to higher radii.
A similar effect can be seen in FP models, see Einsel (1996), Figures 40 and 42.

\section{Isolated two-mass models}

\begin{table*}
\begin{minipage}{140mm}
\begin{tabular}{l|l|l|l|l|l|l||l|l|l|l|l|l|l}
\hline
\hline
Model & $W_0$ & $\omega_0$ & N & $\mu$ & q & $t_\mathrm{cc}/t_\mathrm{rh}$ & Model & $W_0$ & $\omega_0$ & N & $\mu$ & q & $t_\mathrm{cc}/t_\mathrm{rh}$ \\
\hline
TM1a & 3 & 0.0 & 5K & 1.25 & 0.1 & 9.07 & TM2a & 3 & 0.0 & 5K & 2 & 0.1 & 5.38 \\
TM1b & 3 & 0.3 & 5K & 1.25 & 0.1 & 9.77 & TM2b & 3 & 0.3 & 5K & 2 & 0.1 & 5.49 \\
TM1b & 3 & 0.6 & 5K & 1.25 & 0.1 & 9.92 & TM2c & 3 & 0.6 & 5K & 2 & 0.1 & 5.93 \\
\hline
TM3a & 3 & 0.0 & 5K & 5 & 0.1 & 1.30 & TM4a & 3 & 0.0 & 5K & 10 & 0.1 & 0.93 \\
TM3b & 3 & 0.3 & 5K & 5 & 0.1 & 1.43 & TM4b & 3 & 0.3 & 5K & 10 & 0.1 & 0.71 \\
TM3c & 3 & 0.6 & 5K & 5 & 0.1 & 1.39 & TM4c & 3 & 0.6 & 5K & 10 & 0.1 & 0.80 \\
\hline
\hline
TM5a & 6 & 0.0 & 5K & 1.25 & 0.1 & 6.94 & TM6a & 6 & 0.0 & 5K & 2 & 0.1 & 3.86 \\
TM5b & 6 & 0.3 & 5K & 1.25 & 0.1 & 7.25 & TM6b & 6 & 0.3 & 5K & 2 & 0.1 & 4.09 \\
TM5b & 6 & 0.6 & 5K & 1.25 & 0.1 & 7.08 & TM6c & 6 & 0.6 & 5K & 2 & 0.1 & 4.16 \\
\hline
TM7a & 6 & 0.0 & 5K & 5 & 0.1 & 0.84 & TM8a & 6 & 0.0 & 5K & 10 & 0.1 & 0.28 \\
TM7b & 6 & 0.3 & 5K & 5 & 0.1 & 0.89 & TM8b & 6 & 0.3 & 5K & 10 & 0.1 & 0.31 \\
TM7c & 6 & 0.6 & 5K & 5 & 0.1 & 1.21 & TM8c & 6 & 0.6 & 5K & 10 & 0.1 & 0.51 \\
\hline
\hline
TM9a & 3 & 0.0 & 32K & 25 & 0.1 & 0.27 & TM10a & 3 & 0.0 & 32K & 50 & 0.1 & 0.26 \\
TM9b & 3 & 0.3 & 32K & 25 & 0.1 & 0.37 & TM10b & 3 & 0.3 & 32K & 50 & 0.1 & 0.22 \\
TM9c & 3 & 0.6 & 32K & 25 & 0.1 & 0.31 & TM10c & 3 & 0.6 & 32K & 50 & 0.1 & - \\
\hline
TM11a & 6 & 0.0 & 32K & 25 & 0.1 & 0.10 & TM12a & 6 & 0.0 & 32K & 50 & 0.1 & 0.06 \\
TM11b & 6 & 0.3 & 32K & 25 & 0.1 & 0.12 & TM12b & 6 & 0.3 & 32K & 50 & 0.1 & 0.07 \\
TM11c & 6 & 0.6 & 32K & 25 & 0.1 & 0.13 & TM12c & 6 & 0.6 & 32K & 50 & 0.1 & 0.08 \\
\hline
\end{tabular}
\end{minipage}
\caption{The initial models with two mass components.}
\label{TM}
\end{table*}

Realistic globular clusters have a mass spectrum. However, the main effects
of a mass spectrum on the dynamics of globular clusters can
already be seen in idealized models composed out of only two mass components.
We study such models in this section.
Table \ref{TM} summarizes our initial models with two mass components.
All these models are done with {\sc nbody6++}. We would like to stress that they are isolated, 
and no averaging was used. 
The total mass fraction of heavy stars $q$ was fixed 
to the value $q=0.1$ as for the largest part of the parameter space covered in Khalisi (2002) 
and Khalisi et al. (2006). As he notes, this choice 
was inspired by a result of Inagaki \& Wiyanto (1984) that a 
cluster evolves fastest for $q=0.1$.
Following the work of Khalisi, we varied the stellar mass ratio $\mu$.
The models of series TM9 and TM10 with high particle numbers ($N$=32K) were computed 
(using 32-128 processors) at the IBM Regatta p690+ supercomputer ``Jump'' at the Research 
Center J\"ulich, consisting of a total of 1312 Power4+ processors running at a frequency of 
1.7 GHz, whereas the models 
of series TM11 and TM12 (also $N$=32K) were computed with the PC Beowulf 
cluster ``Hydra'' of the ARI, consisting of 10 Dual P4 with 2.2 GHz. 
Each of the core collapse times given in Table \ref{TM} is the arithmetic
mean of three times which were determined for each run: When the core radius had its first 
sharp minimum, the minimum potential in the star cluster reached its first sharp minimum
and the maximum density in the core reached its first sharp maximum.

\begin{figure}
\centering
\includegraphics[width=0.47\textwidth]{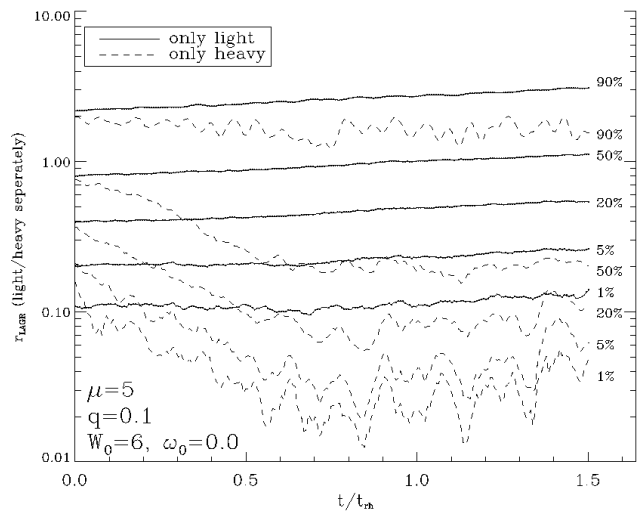}
\caption{Time evolution of the Lagrangian radii of the model TM7a, computed seperately
for the light masses, neglecting the heavy masses and for the heavy masses,
neglecting the light masses. In particular, this Figure shows the radii, in which 
1\%, ..., 90\% of the light/heavy masses are contained.}
\label{fig:lagrlight}
\end{figure}

Figure \ref{fig:lagrlight} shows the time evolution of the Lagrangian radii for the
model TM7a, computed seperately for the light/heavy component.
While the orbits of the light stars expand, the heavy stars sink to the center.
As can be seen, only the heavy masses are involved in the core collapse.
For the non-rotating case with $q=0.1$, the transition between the Spitzer-stable 
and Spitzer-unstable regimes lies indeed between $\mu=1.25$ and $\mu=2$, as we checked 
in our data. For details on the
Spitzer instability, see the original work of Spitzer (1969) or the review
in Khalisi (2002). The empirical criterion by Watters et al. (2000), which is based on 
Monte-Carlo simulations, also yields a transition occuring between the above values 
of $\mu$. 

\begin{figure*}
\centering
\includegraphics[width=0.9\textwidth]{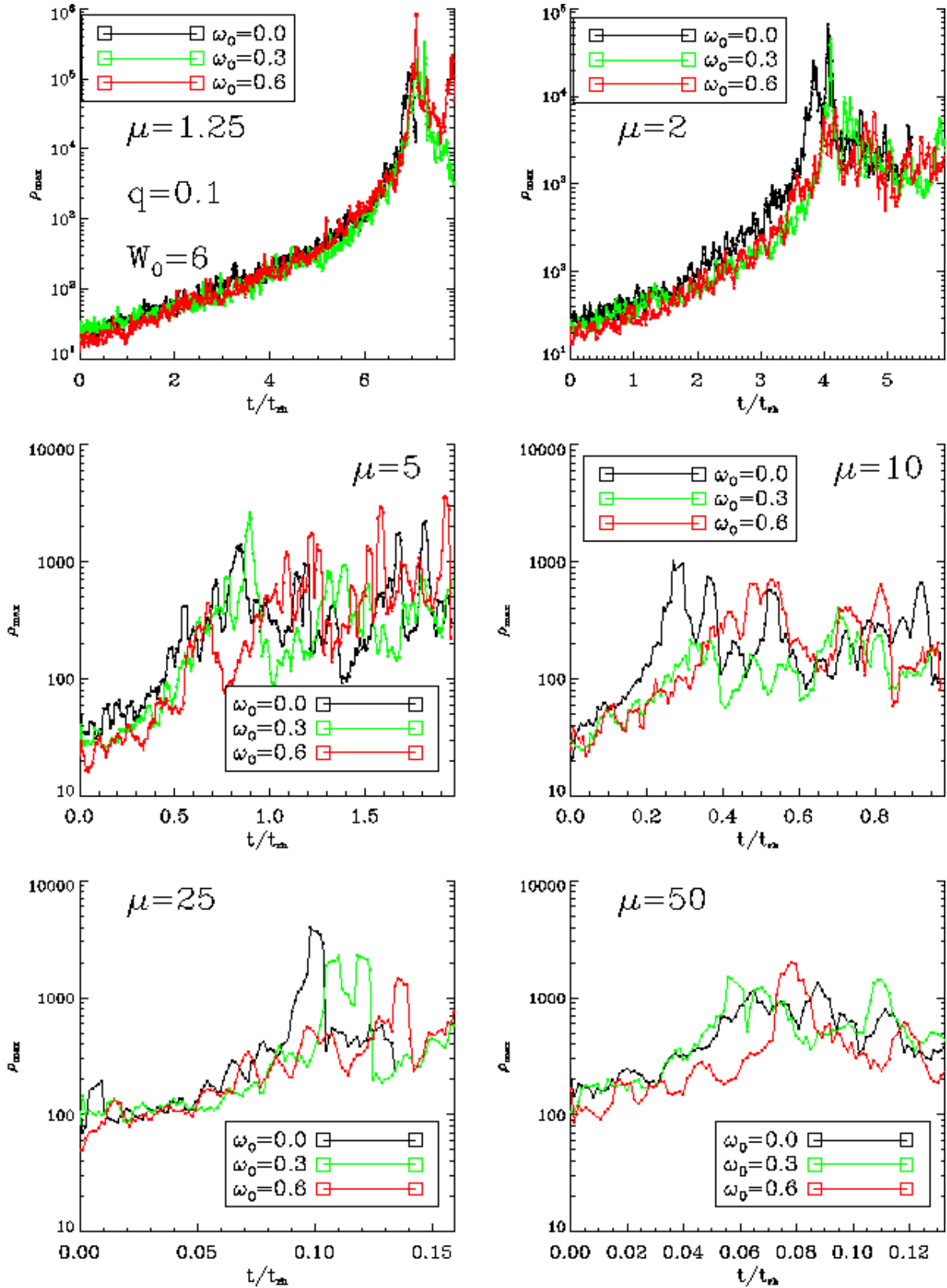}
\caption{Time evolution of the maximal density in the core for the models of series
TM5 -- TM8 and TM11 -- TM12}
\label{fig:rho2mass}
\end{figure*}

\begin{figure}
\centering
\includegraphics[width=0.48\textwidth]{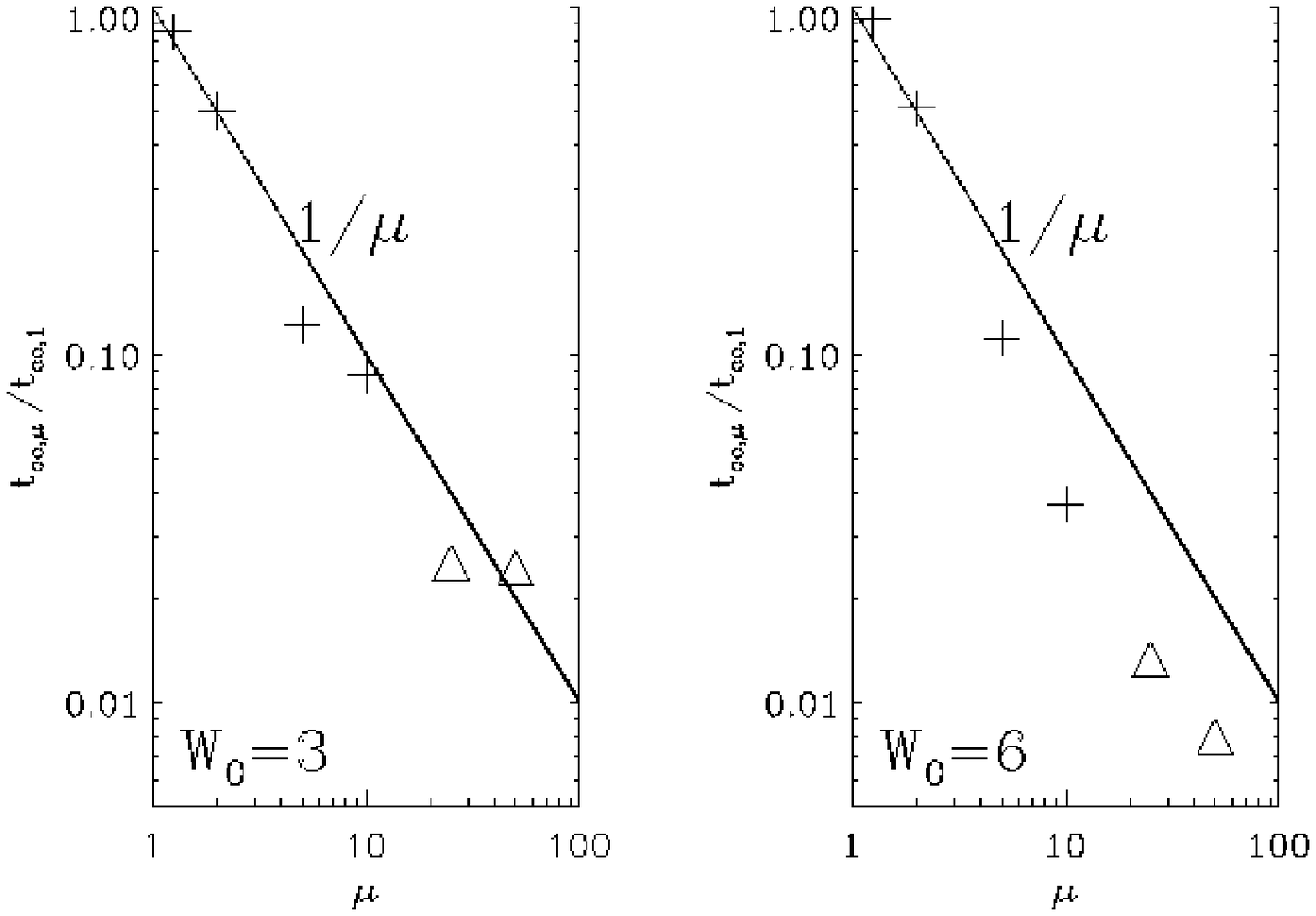}
\caption{Acceleration factor of the core collapse due to mass segregation. The positions
of the crosses (N=5000) and triangles (N=32000)
are calculated from values in Table \ref{TM} for the non-rotating models and normalized to
values in Table \ref{EM12} for the equal-mass models ($\mu=1$).}
\label{fig:accfactor}
\end{figure}

Figure \ref{fig:rho2mass} shows the evolution of the maximal density in the core for the models of 
series TM5 -- TM8 and TM11 -- TM12. What can immediately be seen in Figure \ref{fig:rho2mass} is the 
acceleration of the core collapse due to mass segregation. The higher the stellar
mass ratio $\mu$, the shorter is the core collapse time.

Figure \ref{fig:accfactor} shows the acceleration factor of the core collapse
due to mass segregation. For this plot we used the
equal-mass core collapse times $t_{cc,1}$ of the non-rotating 
isolated models EM1a and EM2a (see Section 3) and the
core collapse times $t_{cc,\mu}$ from the non-rotating models ``a'' 
of series TM1-TM12. 
Khalisi and Spurzem (2001) confirmed the relation

\begin{equation}
t_\mathrm{cc,\mu}\propto t_\mathrm{cc,1}/\mu, \label{eq:accfactoremil}
\end{equation}

\noindent
which is derived from the relevant time scales and also shown in Figure \ref{fig:accfactor}. 
In the limit of small $\mu$ the curve approaches the Spitzer-stable regime while in the
limit of large $\mu$ $N$-dependant effects play a role, which can be seen in more
detail in the study by Khalisi et al. (2006) (their Figure 6).  

However, in Figure \ref{fig:rho2mass} the acceleration of the core collapse due
to rotation is not detectable. Rather, by also taking into account the
core collapse times in columns 7 and 14 of Table \ref{TM}, it seems, that in most models, 
there is an opposite effect, i.e. the rotating models have a delayed core 
collapse. 

\begin{figure}
\centering
\includegraphics[width=0.47\textwidth]{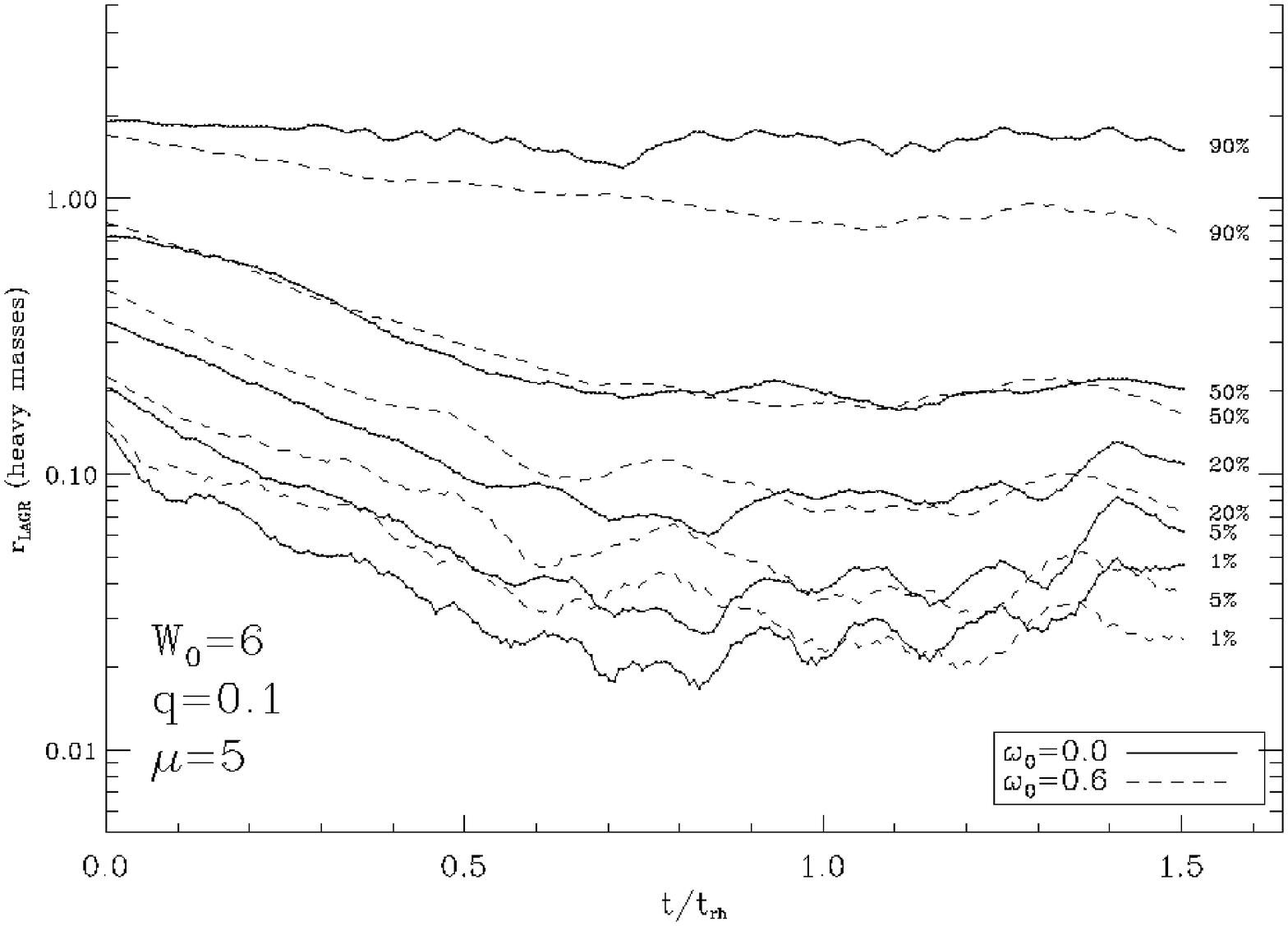}
\caption{Time evolution of the Lagrangian radii of the models TM7a and
TM7c, computed only for the heavy masses, neglecting the light masses.}
\label{fig:lagrheavy}
\end{figure}

Figure \ref{fig:lagrheavy} shows the time evolution of the Lagrangian radii for
the models TM7a and TM7c, computed only for the heavy masses.
A comparison of the curves for the models with different $\omega_0$ in 
Figure \ref{fig:lagrheavy} reveals that the inner ($\lesssim 50\%$) Lagrangian radii 
for the model ($\omega_0=0.6$) decrease somewhat slower 
than those of the non-rotating model ($\omega_0=0.0$). This points towards a 
slowdown of the mass segregation process induced by rotation.

\begin{figure}
\centering
\includegraphics[angle=90,width=0.48\textwidth]{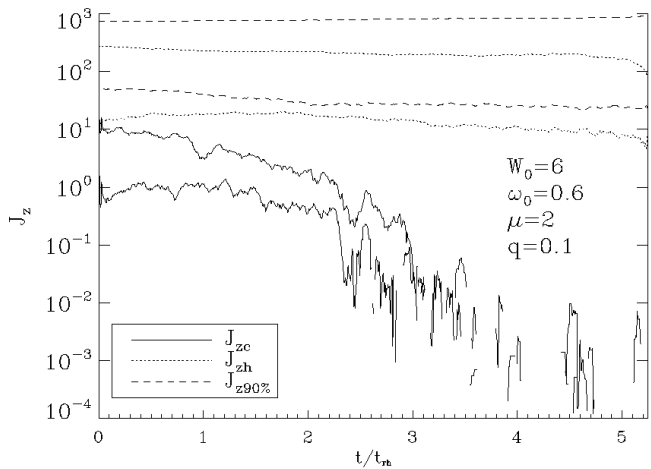}
\caption{Time evolution of the $z$-component of angular momentum contained
within the core, the half-mass radius and the 90\% Lagrangian radius, separately
for the two mass components (for each linestyle: upper curve: light, lower curve: heavy)
for the model TM6c. The interruptions in the black curves stem from the sign
change of the plotted quantity.}
\label{fig:jzlightheavy}
\end{figure}

Figure \ref{fig:jzlightheavy} illustrates the time evolution of the $z$-component of angular 
momentum contained within the core, the half-mass radius and the 90\% Lagrangian 
radius for the two mass species seperately for the isolated model TM6c. 
For each color, the upper curve corresponds to the light masses, while the lower 
curve corresponds to the heavy masses.
Note that the $z$-component of the total angular momentum remains approximately 
constant, since the model is isolated. 
The core loses angular momentum to the outer parts of the cluster. 
Kim et al. point out, that angular momentum is transferred from
the high masses to the low masses. We cannot prove this trend in 
Figure \ref{fig:jzlightheavy}. As we consider the angular momentum 
contained in the core, it can be seen that the high masses 
and the low masses as well lose angular momentum. 
The angular momentum loss of the heavy masses is a bit slower than 
that of the light masses which is still in accordance with K04 (see their Figure 11).

At last, we would like to mention that we checked the pairing of the dynamically 
formed binaries in our two-mass models. In the Spitzer-stable
regime ($\mu=1.25$) all pairings occur: Heavy--heavy, light--light and
heavy--light. The pairings change from time to time. In the Spitzer-unstable 
regime of our parameter space ($\mu=2 - 50$), we noticed
(with a few exceptions) only heavy--heavy binaries.

\section{Discussion}

The main aim of this paper was to investigate the effect of an overall 
(differential) rotation on the dynamical evolution of
globular clusters with direct $N$-body models. 
We would like to briefly summarize our results first and then tackle 
a comparison with results obtained by others.

\subsection{Summary}

The difference between rotating and non-rotating equal-mass $N$-body models of
globular clusters is that an overall (differential) rotation accelerates their dynamical 
evolution. For instance, the core collapse of equal-mass models is accelerated for the cases 
with rotation as compared with the non-rotating models. However, this acceleration of the 
dynamical evolution induced by rotation cannot be seen in our isolated two-mass $N$-body 
models.

As can be seen see in Table 1, the core collapse times are larger for the models with
King parameter $W_0=3$ than for those with $W_0=6$, which are
more concentrated. This is
a well-known fact, which is noted, for instance, in Quinlan
(1996) or G\"urkan et al. (2004). 
For the models EM1a-c with $W_0=3$, the acceleration of the core collapse 
due to rotation is not visible within the estimated error margins of
the measurement. Therefore, runs with higher
particle numbers are required, or, an averaging over several runs
with different initial configurations of positions and
velocities, which has been done for the models EM2a-d with $W_0=6$ in order
to damp statistical fluctuations. 

In the tidally limited models, we observe rapid mass loss across the tidal boundary, which is 
stronger for the faster rotating models. A tidal field further accelerates the core collapse, 
since it provides an additional mechanism for the transport and removal of energy and 
angular momentum through escaping stars, thus enhancing the effects of the
gravothermal and gravogyro instabilities.

In the case of tidally limited $N$-body models within the tidal approximation (see Appendix A), 
the sense of rotation plays an imporant
role in the escape process. Models, where most stars are in retrograde 
orbits (as compared with the sense of rotation of the star cluster around the
galaxy), have a significantly lower escape rate due to the presence of a ``third
integral'', cf. Fukushige \& Heggie (2000) and Appendix B.

The gravogyro instability predicted by Inagaki \& Hachisu (1978) and Hachisu (1979, 1982), 
as it becomes manifest in a decrease of
the $z$-component of angular momentum combined with an increase in
the angular speed, was found in both isolated and tidally limited 
equal-mass $N$-body models.

In the models with two mass components, mass segregation takes place, which again 
accelerates the dynamical evolution of the globular cluster and results in a faster core 
collapse, the higher the stellar mass ratio $\mu$ is. Our results are in fair agreement
with results in Khalisi et al. (2006).

However, if both mass segregation and rotation compete in the acceleration 
of the dynamical evolution of the globular cluster, the trend that rotation accelerates 
the dynamical evolution of the stellar system, is no more visible in isolated models. 
Rather, in most models, there seems to be an opposite effect: The faster 
rotating isolated models have a delayed core collapse as compared with
the non-rotating models. A possible explanation is that the rotation slows
down the mass segregation process. However, the statistical quality of
our results is limited.

\subsection{Comparisons with other $N$-body models}

A few $N$-body models of rotating globular clusters have
been presented in chapter 4 of K03. This chapter is a 
comparative study between $N$-body simulations and FP model results;
see also the recent work by Kim et al. (2006) [hereafter: K06].
Kim's initial models are the same rotating King models as they are described
in Section 2. He used a version of Aarseth's original code {\sc nbody6} 
modified ``to mimic the tidal environment of the clusters modeled with the
2D FP equation.'' Namely, he applied a tidal energy cutoff
to his $N$-body models (see Appendix B). With the same definition
of relaxation time, the half-mass times of his and our $N$-body models differ significantly,
as can be seen in Figure \ref{fig:massloss}. The same is valid for a comparison
of the half-mass times of our non-rotating $N$-body model and the corresponding 
$N$-body model by Heggie (also
with energy cutoff). The reason for these rather large differences is the 
different treatment of the tidal field.
The core collapse times from our $N$-body models agree well with those
of Kim's $N$-body models. Our results are also consistent with $N$-body 
simulations presented in Ardi et al. (2005, 2006) if the same definition of the half-mass
relaxation time is used. 

\subsection{Comparison with FP models}

For a comparative study of $N$-body models of rotating star clusters with
FP models we refer to chapter 4 of K03 and the recent work K06.
While the time step in $N$-body simulations is a fraction of the orbital time,
it is a fraction of relaxation  time in the FP codes used
in ES99 and K02, K03, K04 and K06. Therefore, the proper definition of relaxation
time is crucial for a comparison of $N$-body and FP models.
With equation (\ref{eq:trh}) as definition of the half-mass relaxation time, the core collapse times
in our equal-mass $N$-body models (i.e. the models of series EM1--EM4)
are shorter than the core collapse times in the tidally limited FP models in ES99 
(see their Figure 2) and the isolated FP models in K02 (see their Figure 3).
A similar trend has been found in K03 and K06.

The different implementations of a tidal boundary in $N$-body and FP models
and the absence of tidal forces in FP models  have a significant influence on the
escape rates (see Section 4.2 and Figure \ref{fig:massloss}).

In the mdels with two mass components, a difference
to FP models of K04 occurs: In our isolated models, we do not see the effect 
of acceleration of the core collapse due to rotation any more.

\subsection{Future work}

Several questions remained unanswered --
the endeavor to answer some of the following questions may 
be the starting point for future investigations:
Does the suspected slow-down of the mass segregation process 
induced by rotation also occur in tidally limited models with
two mass components? How evolves the average angular speed of the different 
mass components, taken seperately? We need a more intensive $N$-body study 
of tidally limited rotating systems with a mass spectrum
and a better statistical quality, i.e. higher particle numbers or
extensive averaging. How do stellar evolution, disk shocking, 
primordial binaries or a  central black hole influence the dynamical 
evolution of rotating globular clusters?

\section*{Acknowledgments}

We are indebted to Sverre Aarseth for the provision of {\sc nbody6}.
Furthermore, the authors would like to thank Eunhyeuk Kim and Douglas Heggie
for providing some of their $N$-body models with energy cutoff
for comparison, Peter Berczik for his help with the determination of the critical $\omega_0$
for which isolated rotating King models become unstable to the formation of a bar and 
Marc Freitag for his helpful comments. AE was supported by the International Max Planck Research School for Astronomy and Cosmic Physics (IMPRS) at the University of Heidelberg and would
like to thank Jonathan M. B. Downing for a discussion. We acknowledge 
computing time at the supercomputer ``Jump'' at the NIC J\"ulich, Germany. The PC cluster
``Hydra'' at the ARI (funded by the SFB 439, University of Heidelberg)
was also used.

\appendix

\section{The tidal approximation}

\begin{figure}
\includegraphics[angle=-90,width=0.48\textwidth]{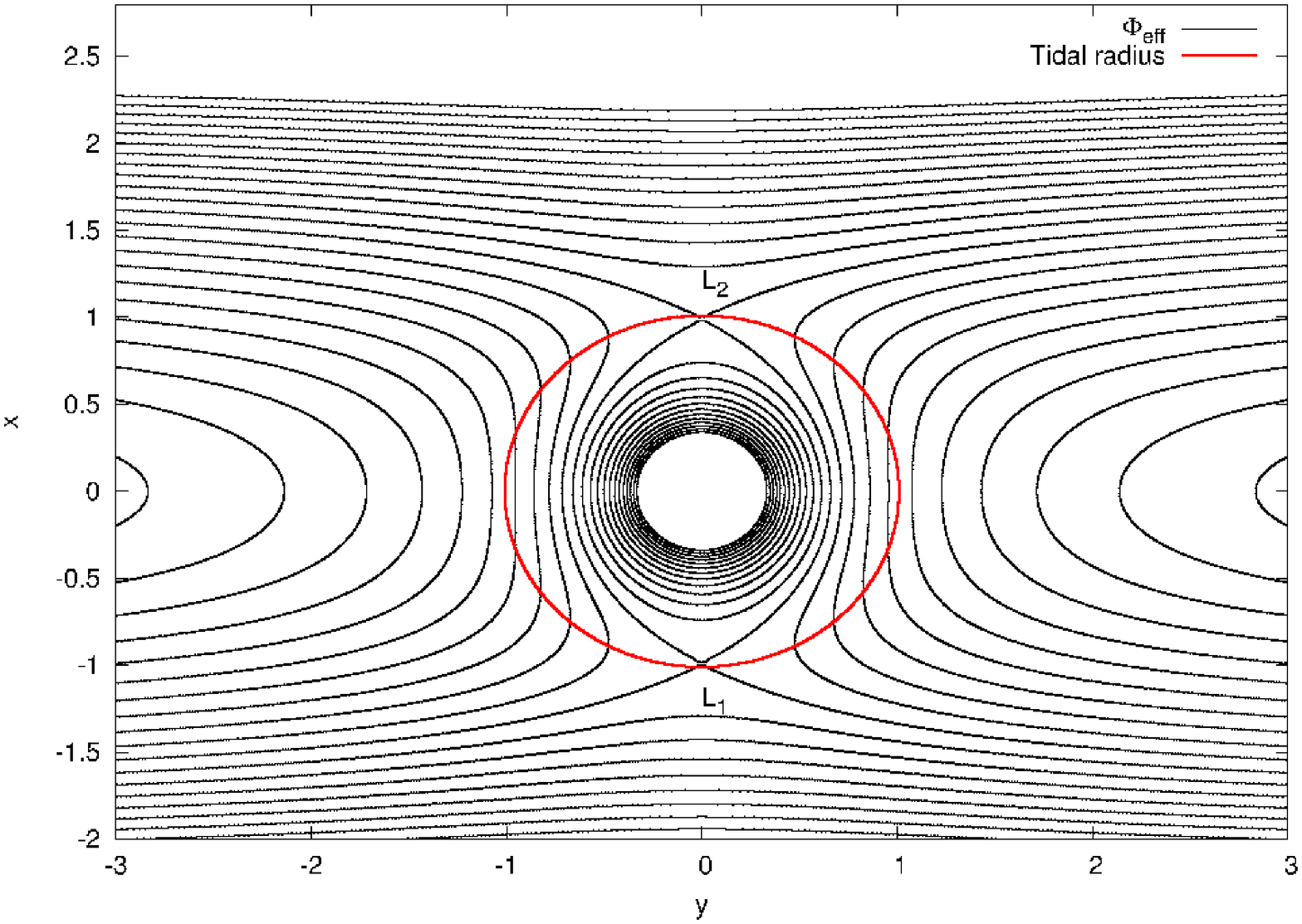} 
\caption{Equipotential lines in the z=0 plane for the effective potential (\ref{eq:phieff}) in
the case $\kappa^2 \simeq 1.8\,\omega_0^2$, $\omega_0=G=1$. For the star cluster we assumed a
Plummer potential, for which we defined a ``concentration'' $c=\log_{10}(r_t/r_{Pl})=1$ and a 
total mass $M\simeq 2.2$. In these units, we have $r_t = 1$ for the tidal radius and 
$C_L\simeq -3.3$ for the critical Jacobi integral.} 
\label{fig:equipot}
\end{figure}

The mean eccentricity of the known 3D globular cluster orbits in the halo of the 
Milky Way is rather high, see, e.g., Dinescu et al. (1999) or Allen et al. (2006)
for a more recent work.
Therefore, a realistic tidal field acting on a globular cluster in the Galactic halo
should be time-dependant. However, in {\sc nbody6} and {\sc nbody6++}, the 
star cluster is assumed to move around the Galactic center on a circular orbit,
which makes it possible to use the epicyclic approximation, 
cf. Binney \& Tremaine (1987). 
The corresponding tidal field is steady.
Its implementation in the $N$-body code is 
as follows: As in the restricted three-body problem, one applies a coordinate 
transformation to a rotating coordinate system with the angular velocity $\omega_0$, 
in which both the star cluster center and the Galactic center (i.e., the primaries) are at rest. 
Its origin is the star cluster center, sitting in the minimum of the effective Galactic potential. 
The x-axis points away from the Galactic center. The y-axis points in the 
direction of the rotation of the star cluster around the Galactic center. The z-axis lies 
perpendicular to the orbital plane
and points towards the Galactic North pole. 
Through this transformation into a frame of reference, in which the
tidal potential is static, centrifugal and Coriolis forces appear
according to classical mechanics. In addition, tidal terms enter
the modified equations of motion in the star cluster region. These can be 
derived from an effective potential in the epicyclic approximation:

\begin{eqnarray}
\Phi_\mathrm{eff}(x,y,z)&=&\Phi_{\mathrm{cl}}(x,y,z) + \frac{1}{2}\mu^2 x^2 
+ \frac{1}{2}\nu^2 z^2 \nonumber \\
&& + \ \mathcal{O}(xz^2)  \label{eq:phieff}
+ \mathrm{const}  
\end{eqnarray}

\noindent
where the coordinates are relative to the star cluster center and
$\Phi_{\mathrm{cl}}$ is the star cluster potential. Since the expansion is about the
minimum of $\Phi_{\rm eff}$, where the first partial derivatives vanish,
there are no first-degree terms in the Taylor expansion. The term $\propto xz$
vanishes because $\Phi_{\rm eff}$ is symmetric in $z$. The term $\propto x^2$
arises from a combination of centrifugal and tidal forces. For an illustration
using a Plummer potential for the star cluster, see
Figure \ref{fig:equipot}, which also shows the location of the effective potential's 
nearest equilibrium points with respect to the star cluster center, which are
covered by the approximation: The Lagrangian points $L_1$ and $L_2$, where 
$\nabla \Phi_{\rm eff} = 0$.
They are saddle points of the effective potential, i.e. the Hessian is neither positive
nor negative definite.
In terms of Oort's constants $A$ and $B$, we have in the solar neighborhood

\vspace{-0.3cm}

\begin{eqnarray}
\vec{\omega}_0 &=& (0,0,\omega_0) = (0, 0, A-B), \\
\kappa^2 &=& -4B(A-B), \\
\mu^2 &=& \kappa^2 - 4\omega_0^2 = -4A(A-B) < 0, \\
\nu^2 &=& 4\pi G\rho_g+2(A^2-B^2) > 0, \label{eq:vertfreq}
\end{eqnarray}

\noindent
where $\kappa$ and $\nu$ are the epicyclic and vertical frequencies,
respectively. The ratio $\kappa^2/\omega_0^2\simeq 1.8$ calculated from
Oort's constants depends in the general case 
on the density profile of the Galaxy. The vertical frequency $\nu$ can be derived from
the Poisson equation for an axisymmetric system, see Oort (1965),
and $\rho_g$ is the local Galactic density, which contributes to the
dominant first term in (\ref{eq:vertfreq}).\footnote{Actually, such a treatment
is only valid in the solar neighborhood, i.e. in the Galactic disk.}
The equations of motion in 
the rotating reference frame are then

\begin{equation}
\ddot{\vec{x}}= -\nabla\Phi_\mathrm{eff} - 2(\vec{\omega}_0\times\dot{\vec{x}}),
\end{equation}

\noindent
where the last term on the right-hand side represents the Coriolis forces,
which cannot be derived as the usual gradient of a potential, since 
they are velocity-dependant. After a little bit of vector analysis, the equations
of motion read 

\vspace{-0.3cm}

\begin{eqnarray}
\ddot{x}&=&f_x + 2(A-B) \dot{y} + 4A(A-B) x \label{eq:eqm1} \\
\ddot{y}&=&f_y - 2(A-B)\dot{x} \label{eq:eqm2} \\
\ddot{z}&=&f_z - \left[4\pi G\rho_g+2(A^2-B^2)\right] z, \label{eq:eqm3} 
\end{eqnarray}

\noindent
where $(f_x, f_y, f_z) = -\nabla\Phi_{\mathrm{cl}}$ is the force vector from the other cluster member stars.

\section{Escape criteria}

\begin{table}
\begin{minipage}{\textwidth}
\begin{tabular}{l|l}
\hline
\hline
Isolated system & Tidally limited system \\
\hline
1. $E_\star > 0$ & 1. $ C_\star > C_L$ \\
2. $\vert\vec{r}_\star-\vec{r}_d\vert > r_\mathrm{crit} = 20\cdot r_V$ & 2. $\vert\vec{r}_\star-\vec{r}_d\vert > 2\cdot r_t$ \\
\hline
\end{tabular}
\end{minipage}
\caption{Escape criteria in  {\sc nbody6}  and {\sc nbody6++}.}
\label{escape}
\end{table}

\medskip

\begin{figure}
\includegraphics[width=0.48\textwidth]{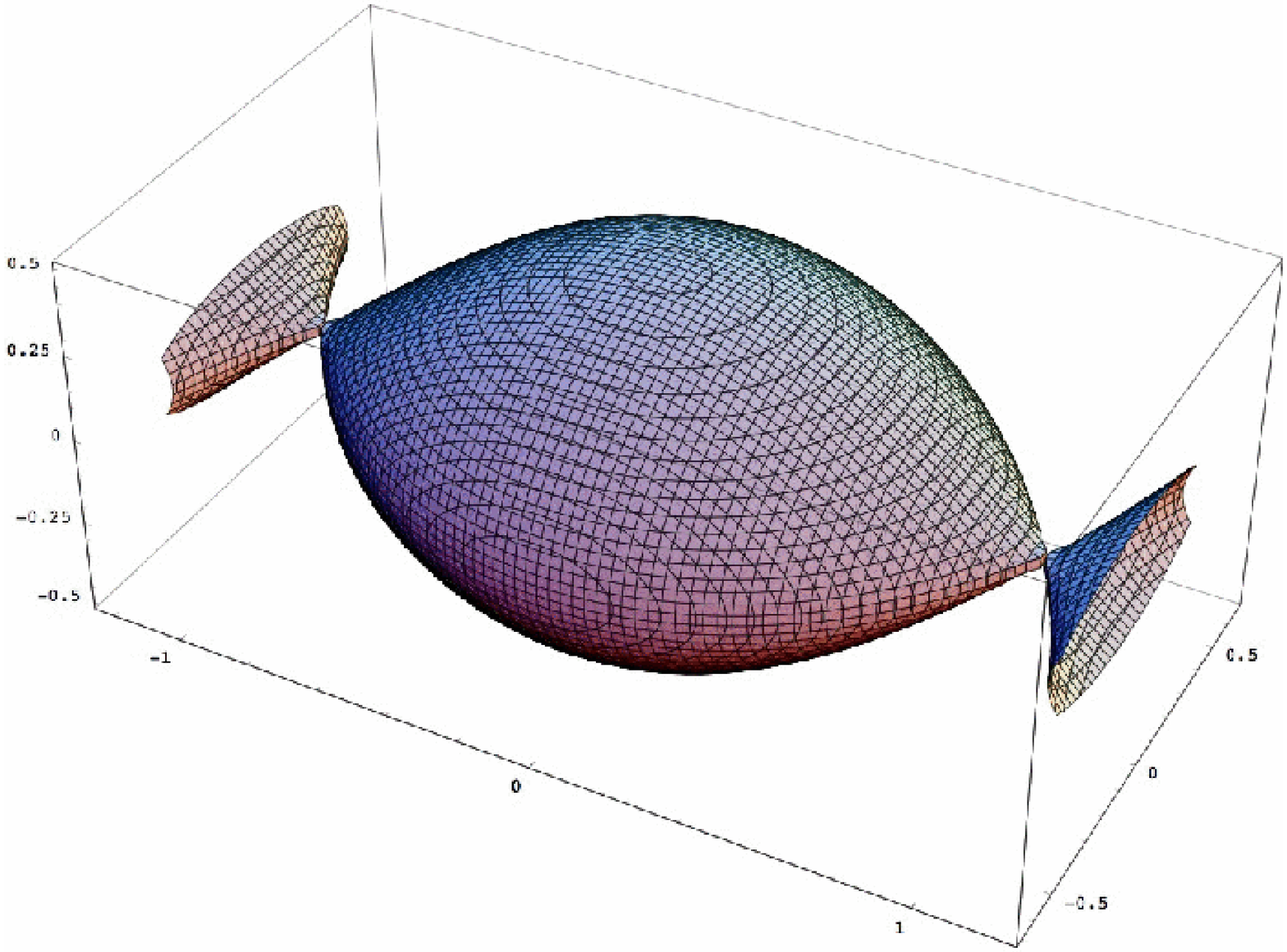} 
\caption{The critical equipotential surface at $\Phi_{\rm eff}=C_L$ with the same conditions 
as in Figure \ref{fig:equipot}. We additionally assumed $\nu^2 \simeq 8.5 \, \omega_0^2$.
The $z$-axis is pointing upwards.} 
\label{fig:hillsurf}
\end{figure}

In {\sc nbody6}  and {\sc nbody6++}, the escape criteria
shown in Table \ref{escape} are implemented. $E_\star$  is the energy of a star,
$C_\star$ its Jacobi integral, $\vec{r}_\star$  its position vector,
$\vec{r}_d$ the position of the star cluster's density center and $r_V$ is
the virial radius. 
In both the isolated and tidally limited cases, there are a necessary and a sufficient
criterion for escape implemented in {\sc nbody6}  and {\sc nbody6++}. Since the
sufficient criterion is always related to a critical radius, we refer to 
these escape criteria as \emph{radius cutoffs}.

The tidal radius is given by    

\begin{equation}
r_t=\left[\frac{GM}{4A(A-B)}\right]^{1/3}=\left(\frac{M}{3M_g}\right)^{1/3} R_0
\end{equation}

\noindent
according to King (1962), where $G$ is the gravitational constant, $M$ is the total mass
of the star cluster, $A$ and $B$ are Oort's constants $M_g$ is the mass of the Galaxy
and $R_0$ is the Galactocentric radius of the star cluster's orbit around the Galactic center.
The critical Jacobi integral is given by the value of the effective potential (\ref{eq:phieff}) 
at the Lagrangian points $L_1$ and $L_2$,

\begin{equation}
C_L = -\frac{3GM}{2r_t} = -\frac{3}{2}\left[G^2 M^2 4A(A-B)\right]^{1/3}
\end{equation}

\noindent
according to Wielen (1972). It is conserved in the rotating reference frame,
which is centered on the star cluster center. For $C_\star>C_L$, the equipotential surfaces in 
Figure A1 are open and form channels of escape, through which orbits can
leak out. The last closed surface is almond-shaped (see Figure \ref{fig:hillsurf}).

Takahashi \& Portegies Zwart (1998) note that ``In an isotropic Fokker-Planck
model, one has no choice but to use the energy as a criterion for escape.''
What they refer to is the so-called \emph{energy cutoff} in contrast to 
the \emph{radius cutoff} of the $N$-body codes. Stars with

\begin{equation}
E>E_t= \Phi(r_t) = -\frac{GM}{r_t} \label{eq:etid}
\end{equation}

\noindent
are removed from the system, where $E_t$ is the tidal energy, i.e. the
potential at the tidal radius. This criterion has already been used
by Chernoff \& Weinberg (1990). Stars, whose
energy exceeds the tidal energy by a small amount, are able to cross 
the tidal radius (if they are on radial orbits). In the energy cutoff picture
within the FP approximation, the lifetime  of a star cluster scales exactly with 
relaxation time. However, such an approach may simplify the physics: 
Once the star's energy is higher than the tidal energy it is called a
``potential escaper'': It has not yet escaped but still needs a time of the order 
of the crossing time to reach the tidal radius. Immediate removal of stars which fullfill 
the energy criterion is only reasonable if the orbital timescale is 
negligible compared with the relaxation time. An improved approach, which 
take this dependance on the crossing time scale into account, can be found in the paper by 
Lee \& Ostriker (1987).


In an anisotropic FP code, it is possible to use the more realistic
apocentre criterion by Takahashi et al. (1997) and Takahashi \& 
Portegies Zwart (1998): Stars are removed
from the system, if their apocenter distance $r_a$ calculated according to

\begin{equation}
J^2=2r_a^2\left[E-\Phi(r_a)\right],
\end{equation}

\noindent
exceeds the tidal radius. 



\begin{figure}
\includegraphics[angle=-90,width=0.48\textwidth]{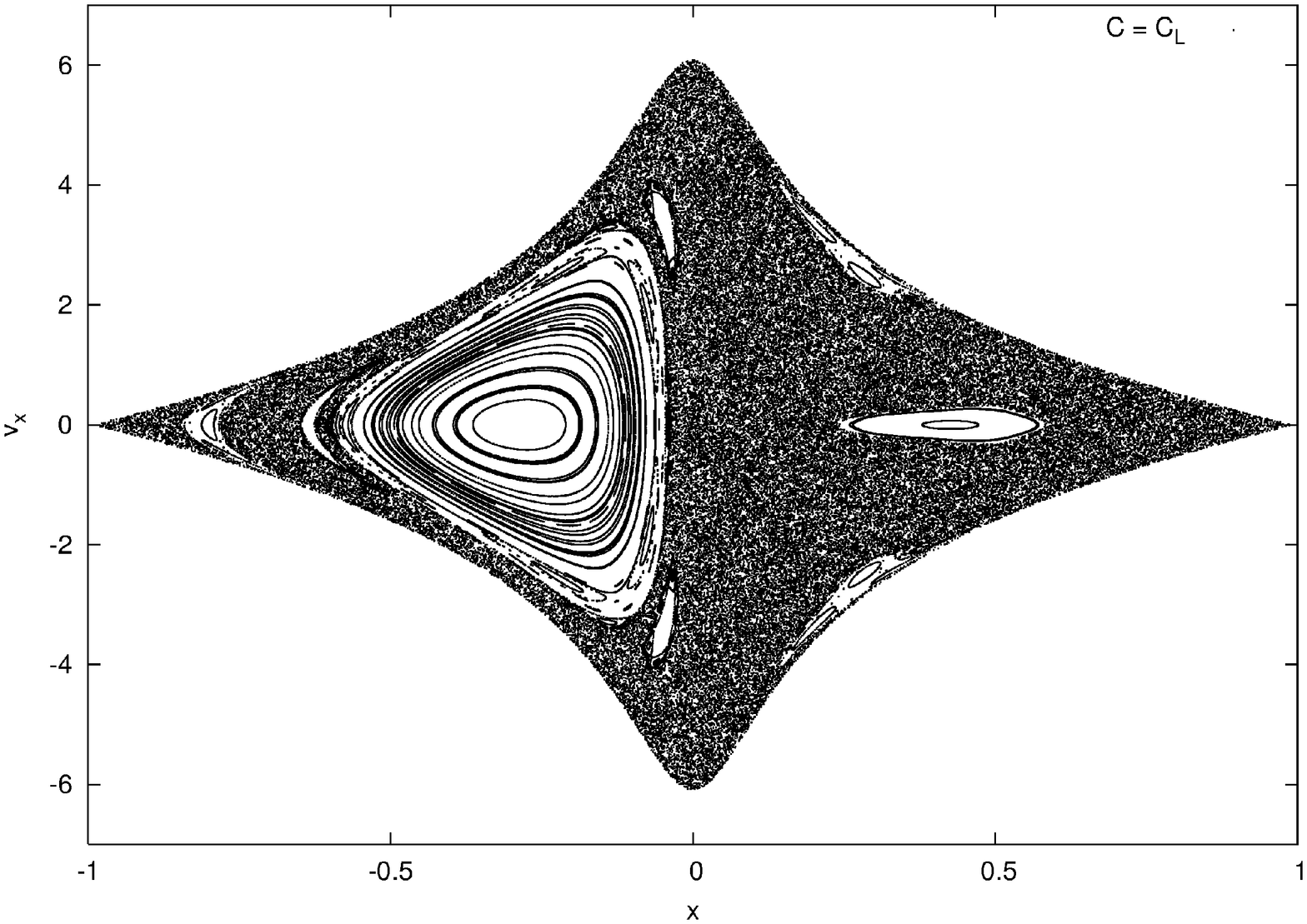} 
\caption{Poincar\'e section of orbits at the critical Jacobi integral $C_L$
for the equations of motion (\ref{eq:eqm1})-(\ref{eq:eqm2}) with the same conditions as in
Figure \ref{fig:equipot} at the moment of crossing $y = 0$ with $\dot{y}>0$. The full 
range of initial conditions is covered.} 
\label{fig:section}
\end{figure}

Last but not least, in the case of a radius cutoff and the tidal approximation, 
the picture is similar to the one sketched above: Once a star's 
Jacobi integral $C_\star$ has exceeded the critical value $C_L$ slightly (i.e. once
the star is a ``potential escaper''), it still needs a certain time to find an opening to one 
of the escape channels
in the equipotential surfaces shown in Figure A1 for the effective potential 
(\ref{eq:phieff}). This can take many crossing times
depending mainly on the excess energy, i.e. we have  $t_e\propto C_L^2(C_\star-C_L)^{-2}$
from an upper  limit on the flux of phase space volume through $L_1 / L_2$ according to 
MacKay (1990) and Fukushige \& Heggie (2000).
In the radius cutoff picture within the tidal approximation, the scaling of the 
lifetime is very subtle and the half-mass time scales as 
$t_{\rm mh}\propto t_{rh}^{3/4}$, see Baumgardt (2001). 

By integrating the equations of motion (\ref{eq:eqm1}) - (\ref{eq:eqm3}) numerically for
orbits at a given Jacobi constant, one can obtain a Poincar\'e section
as shown in Figure \ref{fig:section}. As seen in this Figure, this is a system with divided
phase space: Several ``islands'' of closed invariant 
curves in a stochastic ``sea'' show the existence of an additional conserved quantity other than 
the Jacobian for these orbits. 
The largest island of quasiperiodic orbits in the left half of the surface of section 
corresponds to retrograde orbits, i.e. the stars move around the star cluster in the 
opposite sense to the motion of the star cluster around the galaxy, see Fukushige 
\& Heggie (2000). The ``third
integral'' restricts their accessible phase space and hinders them from escaping,
even if the stars have been scattered above the critical Jacobi constant.
On the other hand, the particles on orbits corresponding to the chaotic domains
in the surface of section bounce back and forth for a certain time in a bounded area, 
called the ``scattering region'', until they escape through one of the exits, which open up
around the Lagrangian points $L_1$ and $L_2$, for stars with a Jacobi
constant, which is higher than the critical value. For an interesting overview of the
physics of chaotic scattering see the papers by Aguirre et al. (2001) and
Aguirre \& Sanju\'an (2003).

\bsp

\label{lastpage}

\end{document}